\documentclass[10pt]{iopart}

\usepackage{t1enc}
\newcommand{\eth}{\mathop{\hbox{\rm \dh}}\nolimits}
\newcommand{\thorn}{\mathop{\hbox{\rm \th}}\nolimits}

\newcommand{\phneg}{\hphantom{-}}

\def \pul {{{\scriptstyle{\frac{1}{2}}}}}
\def \ctvrt {{{\scriptstyle{\frac{1}{4}}}}}
\def \vs {\longleftrightarrow}
\def \mm {\mbox{\quad }}

\def \lvkz {\Bigl(}
\def \pvkz {\Bigr)}

\def \lvhz {\Bigl[}
\def \pvhz {\Bigr]}

\def \BE {\begin{equation}}
\def \EE {\end{equation}}
\def \BEAH {\begin{eqnarray*}}
\def \EEAH {\end{eqnarray*}}
\def \BEA {\begin{eqnarray}}
\def \EEA {\end{eqnarray}}
\def \BDM {\begin{displaymath}}
\def \EDM {\end{displaymath}}

\def \eps {\varepsilon}

\def \na {\alpha}

\def \nb {\beta}
\def \ncb {\bar \beta}
\def \ng {\gamma}
\def \ncg {\bar \gamma}
\def \ne {\varepsilon}

\def \nk {\kappa}
\def \nck {\bar \kappa}
\def \nl {\lambda}

\def \nm {\mu}

\def \nn {\nu}

\def \nr {\rho}
\def \ncr {\bar \rho}
\def \ns {\sigma}
\def \ncs {\bar \sigma}
\def \nt {\tau}
\def \nct {\bar \tau}

\def \d {\delta}

\newcommand{\tA}{\dot A}
\newcommand{\tB}{\dot{B}}
\newcommand{\tC}{\dot{C}}
\newcommand{\tD}{\dot{D}}

\newcommand{\smallsup}[1]{^{\scriptscriptstyle #1}}
\newcommand{\smallsub}[1]{_{\scriptscriptstyle #1}}

\newcommand{\dsu}[2]{\nabla\smallsup{#1\!\dot{#2}}}
\newcommand{\dsd}[2]{\nabla\smallsub{#1\!\dot{#2}}}
\newcommand{\dsud}[2]{\nabla\smallsup{#1}\smallsub{\hphantom{#1}\!\dot{#2}}}
\newcommand{\dsdu}[2]{\nabla\smallsub{#1}\smallsup{\hphantom{#1}\!\dot{#2}}}

\newcommand{\sou}[1]{o\smallsup{#1}}
\newcommand{\sod}[1]{o\smallsub{#1}}
\newcommand{\csou}[1]{{\bar o}\smallsup{\dot{#1}}}
\newcommand{\csod}[1]{{\bar o}\smallsub{\dot{#1}}}
\newcommand{\siu}[1]{\iota\smallsup{#1}}
\newcommand{\sid}[1]{\iota\smallsub{\!#1}}
\newcommand{\csiu}[1]{{\bar \iota}\smallsup{\dot{#1}}}
\newcommand{\csid}[1]{{\bar \iota}\smallsub{\!\dot{#1}}}

\newcommand{\ospu}[2]{{#1}\smallsup{#2}\,}
\newcommand{\ospd}[2]{{#1}\smallsub{#2}\,}

\newcommand{\ospdu}[3]{#1\smallsub{#2}\smallsup{\!\hphantom{#2}\!#3}}

\newcommand{\epsu}[1]{\eps\smallsup{#1}}
\newcommand{\epsd}[1]{\eps\smallsub{#1}}

\usepackage{latexsym}
\newcommand{\qed}{%
\ifmmode \quad\Box
\else
\leavevmode\unskip\penalty9999 \hbox{}\nobreak\hfill
\quad\hbox{$\Box$}%
\fi}

\newcommand{\hfrac}[1]{#1\kern-1pt/\kern -.5pt2}

\newcommand{\authnote}[1]{}

\newcommand{\eqref}[1]{(\ref{#1})}

\newtheorem{proposition}{Proposition}
\newtheorem{theorem}[proposition]{Theorem}
\newtheorem{lemma}[proposition]{Lemma}
\newtheorem{definition}[proposition]{Definition}
\newenvironment{proof}{{\em Proof.}}{\qed\par\medskip}

\begin{document}

\title{All spacetimes with vanishing curvature invariants}
\author{V. Pravda\dag, A. Pravdov\' a\dag, A. Coley\ddag, R. Milson\ddag} 

\address{\dag\ Mathematical Institute, 
Academy of Sciences, \v Zitn\' a 25, 115 67 Prague 1, Czech Republic}
\address{\ddag\ Dept. Mathematics and Statistics, Dalhousie U., 
Halifax, Nova Scotia B3H 3J5, Canada}

\eads{\mailto{pravda@math.cas.cz}, \mailto{pravdova@math.cas.cz},
\mailto{aac@mathstat.dal.ca}, \mailto{milson@mathstat.dal.ca}}

\begin{abstract}
All Lorentzian spacetimes with vanishing 
invariants constructed from the~Riemann tensor and its covariant
derivatives are determined. A subclass of the Kundt spacetimes results and we display the
corresponding metrics in local coordinates. Some
potential applications of these spacetimes are discussed.
\end{abstract}


\section{Introduction}
\authnote{Important: makra1.tex is no longer needed. I imported
the necessary macros into the main file. This should make adding
new macros and sharing
the tex file a bit easier. I also modified some of the macros to
get rid of the trailing blank problem.\\RM-25-04}

A curvature invariant of order $n$ is a scalar obtained by contraction
from a polynomial in the~Riemann tensor and its covariant derivatives
up to the~order $n$. 
In general there are 14 algebraically independent curvature invariants of zeroth order,
the~simplest being the~Ricci scalar. 
Many papers are devoted to studying the properties of the~zeroth order curvature invariants 
(see \cite{Geh} 
-- \cite{CZMJMP02} and references therein) but higher order curvature
invariants remain largely unexplored. Recently it was shown that for
spacetimes in which the~Ricci tensor does not possess a null
eigenvector, an appropriately chosen set of zeroth order curvature
invariants contains all of the information that is present in
the~Riemann tensor \cite{CZMJMP02}. This is certainly not true for
vacuum Petrov type N spacetimes with nonzero expansion or twist, all
of whose zeroth and first order curvature invariants vanish, but for
which there are non-vanishing curvature invariants of the~second order
\cite{typN}; and for some non-flat spacetimes in which all curvature
invariants of all orders vanish \cite{typN}-\cite{Koutras}.

In this paper we shall determine all
Lorentzian spacetimes for which all curvature invariants of all orders are zero.
Indeed, we shall prove the~following:

\begin{theorem}
\label{th:main}
All curvature invariants of all orders vanish if and only if the~following
two conditions are satisfied:
\begin{enumerate}
\item[\rm (A)] The spacetime possesses a non-diverging SFR (shear-free,
geodesic null congruence).
\item[\rm (B)] Relative to the~above null congruence, all curvature
scalars with non-negative boost-weight vanish.
\end{enumerate}
\end{theorem}
The analytic form of the~condition (A), expressed relative to any spin
basis where $\sou{A}$ is aligned with the~null congruence in question,
is simply
\begin{equation}
\nk=\nr=\ns=0\ , \label{eq:Kundt} 
\end{equation}
and the~analytic form of
condition (B) is
\begin{eqnarray}
&& \Psi_0 = \Psi_1 = \Psi_2 = 0\ ,\label{eq:psivanish}\\
&& \Phi_{00} = \Phi_{01} = \Phi_{02} = \Phi_{11} = 0\ , \label{eq:phivanish}\\
&& \Lambda = 0\ . \label{eq:ricvanish}
\end{eqnarray}

Spacetimes that satisfy condition (A) belong to Kundt's class
\cite{kramer,kundt} (also, see Section~\ref{sect:spacetimes} and
\ref{ap:kundt}). Condition (B) implies that the~spacetime is of
Petrov type III, N, or O (see Eq.~\eqref{eq:psivanish}) with
the~Ricci tensor restricted by (\ref{eq:phivanish}) and
(\ref{eq:ricvanish}). (Note: throughout this paper we follow
the~notation of \cite{penrid}; $\Lambda$ is not the~cosmological
constant, it is the~Ricci scalar up to a constant factor.)

The~GHP formalism \cite{penrid} assigns an integer, called the~boost
weight, to curvature scalars and certain connection coefficients and
operators. This is important for this work, and we shall summarize some 
of the~key details of this notion
in the~next section.

The outline for the rest of the paper is as follows.
Section~\ref{sect:sufficient} is devoted to the~proof that the~above
conditions are sufficient for vanishing of curvature invariants.
The~``necessary'' part of Theorem 1 is proved in
Section~\ref{sect:necessary}. The curvature invariants constructed in
this section may be also useful for computer-aided classification of
spacetimes. Kundt's class of spacetimes admits a conveniently
specialized system of coordinates, and so it is possible to classify
and explicitly describe all spacetimes with vanishing curvature
invariants. This is briefly summarized in
Section~\ref{sect:spacetimes}, and some of the details are presented
in \ref{ap:kundt}. We conclude with a discussion.

Perhaps the~best known class of spacetimes with vanishing curvature
invariants are the pp-waves (or plane-fronted gravitational
waves with parallel rays), which are characterized as Ricci-flat
(vacuum) type N spacetimes that admit a covariantly constant null
vector field. The~vanishing of curvature invariants in pp-wave spacetimes has been known
for a long time \cite{jordan}, and the~spacetimes obtained here can perhaps be
regarded as extensions and generalizations of these important
spacetimes. In many applications (e.g., in vacuum pp-wave spacetimes) 
the~resulting exact solutions have a
five-dimensional isometry group acting on three-dimensional null orbits 
(which includes translations in
the~transverse direction along the~wave front) and hence the~solutions
are plane waves. However, the~solutions studied here need not be plane waves,
and are not necessarily vacuum solutions. In particular, non-vacuum
spacetimes with a covariantly constant null vector are often referred
to as generalized pp-wave and typically have no further symmetries
(the~arbitrary function in the~metric is not subject to a further
differential equation, namely Laplace's equation, when the~Ricci
tensor has the~form of null radiation). The pp-wave spacetimes have a
number of remarkable symmetry properties and have been the~subject of
much research \cite{kramer}. For example, the~existence of a homothety
in spacetimes with plane wave symmetry and the~scaling properties of
generally covariant field equations has been used to show that all
generally covariant scalars are constant \cite{scm1} and that metrics
with plane wave symmetry trivially satisfy every system of generally
covariant vacuum field equations {\it except} the~Einstein equations
\cite{torre}.

In addition to pp-waves, presently there are known to be three classes of
metrics with vanishing curvature invariants: the conformally flat pure radiation
spacetime given in \cite{Koutras}; the vacuum Petrov type-N
nonexpanding and nontwisting spacetimes \cite{typN} (this class
contains the~pp-waves); and vacuum Petrov type-III nonexpanding
and nontwisting spacetimes \cite{typIII}. Naturally all of these
spacetimes are subcases of the~class studied here. The spacetimes studied
in \cite{Edgar,Ludwig} also belong to our class.

There are two important applications of the class of spacetimes
obtained in this paper. A knowledge of all Lorentzian spacetimes for
which all of the~curvature invariants constructed from the~Riemann
tensor and its covariant derivatives are zero, which implies that all
covariant two-tensors constructed thus are zero except for the~Ricci
tensor, will be of potential relevance in the~equivalence problem and
the~classification of spacetimes, and may be a useful first step
toward addressing the~important question of when a spacetime can be
uniquely characterized by its curvature invariants. More importantly
perhaps, the~spacetimes obtained in this paper are also of physical
interest. For example, pp-wave spacetimes are exact solutions in
string theory (to all perturbative orders in the~string tension)
\cite{amati,HS} and they are of importance in quantum gravity
\cite{gibbons}. It is likely that all of the~spacetimes for which all
of the~curvature invariants vanish will have similar applications and
it would be worthwhile investigating these metrics further.

Finally, we note that it is possible to generalize
Theorem~\ref{th:main} by including spacetimes with non-vanishing
cosmological constant. The~assumptions regarding the~Weyl
and traceless Ricci tensors remain the~same. Even in this general
case, the~invariants constructed from the~Weyl tensor, the~traceless
Ricci tensor and their arbitrary covariant derivatives vanish.
The~only non-vanishing curvature invariants are order zero curvature
invariants constructed as various polynomials of the~cosmological constant. It must be
noted, however, that there may exist other types of spacetimes with
constant curvature invariants.

\section{Sufficiency of the conditions}
\label{sect:sufficient}

Before tackling the~proof of the main theorem, we make some necessary
definitions and establish a number of auxiliary results. We shall
make use of the~Newmann-Penrose (NP) and the~compacted (GHP) formalisms
\cite{penrid}. 
Throughout we work with a
normalized spin basis $\sou{A}, \siu{A}$, i.e. 
\[ \sod{A}\siu{A} = 1\ . \]
The~corresponding null tetrad is given by
\[
l^{\na} \vs \sou{A}\csou{A}\ , \mm
n^{\na} \vs \siu{A} \csiu{A}\ , \mm
m^{\na} \vs \sou{A} \csiu{A}\ , \mm
{\bar{m}}^{\na} \vs \siu{A} \csou{A}\ 
\] 
with the~only nonzero scalar products 
\[
l_{\na} n^{\na} = - m_{\na} {\bar{m}}^{\na} = 1. \label{nulbaz}
\]
We also recall that 
\BE
\sod{A}\sou{A} =0=\sid{A}\siu{A}\ .\label{oo_ii}
\EE

The~spinorial form of the~Riemann tensor $R_{\na \nb \ng \d} $ is
\BE
\fl
R_{\na \nb \ng \d} \ \ \vs \ospd{X}{ABCD} \epsd{\tA\tB } \epsd{\tC
\tD } + \ospd{\bar X}{\tA \tB \tC \tD } \epsd{AB}
\epsd{CD} +
\ospd{\Phi}{AB \tC \tD } \epsd{\tA \tB }
\epsd{CD} + \ospd{\bar \Phi}{\tA \tB CD} \epsd{AB}
\epsd{\tC \tD } \ , \label{rozpR} 
\EE
where
\BEAH
\ospd{X}{ABCD} = \ospd{\Psi}{ABCD} + \Lambda
(\epsd{AC} \epsd{BD} +\epsd{AD} \epsd{BC} )\ 
\EEAH
and $ \Lambda=R/ 24$, with $R$ the~scalar curvature. 
The~Weyl spinor $\ospd{\Psi}{ABCD}=\ospd{\Psi}{(ABCD)} $ 
is related to the~Weyl
tensor $C_{\na \nb \ng \d} $ by
\BE
C_{\na \nb \ng \d} \ \vs \ \ospd{\Psi}{ABCD} \epsd{\tA \tB } \epsd{\tC
\tD } + \ospd{\bar \Psi}{\tA \tB \tC \tD } \epsd{AB}
\epsd{CD} \ . \label{Wtenz}
\EE
Projections of $ \ospd{\Psi}{ABCD} $
onto the~basis spinors $\sou{A} $, $\siu{A} $
give five complex scalar quantities $ {\Psi}_{0},\ {\Psi}_{1},\
{\Psi}_{2}, \ {\Psi}_{3},\ {\Psi}_{4} $.
The~Ricci spinor 
$ \ospd{\Phi}{AB \tC \tD } = \ospd{\Phi}{(AB)(\tC \tD ) }
= \ospd{\bar \Phi}{AB \tC \tD } $ is connected to the~traceless
Ricci tensor $ S_{\na \nb}= R_{\na \nb} - \ctvrt R g_{\na \nb}$
\BE
\ospd{\Phi}{AB \tA \tB } \vs -\pul S_{ab}\ .
\label{sbsRic}
\EE
The~projections of $\ospd{\Phi}{AB \tA \tB } $
onto the~basis spinors $\sou{A} $, $\siu{A} $
are denoted $\Phi_{00}=\bar{\Phi}_{00}$, $\Phi_{01}=\bar{\Phi}_{10}$, 
$\Phi_{02}=\bar{\Phi}_{20}$, $\Phi_{11}=\bar{\Phi}_{11}$, 
$\Phi_{12}=\bar{\Phi}_{21}$, and $\Phi_{22}=\bar{\Phi}_{22}$.

Eqs.~(\ref{eq:psivanish}) and (\ref{eq:phivanish}) imply
\BEA
\ospd{\Psi}{ABCD}& =& \Psi_4 \sod{A} \sod{B} \sod{C} \sod{D} 
- 4 \Psi_3 \sod{(A} \sod{B} \sod{C} \sid{D)}
\ , \label{Weylspin}\\
\ospd{\Phi}{AB \dot{C} \dot{D}} &=& \Phi_{22} \sod{A} \sod{B} \csod{C} \csod{D} 
- 2 \Phi_{12} \sid{(A} \sod{B)} \csod{C} \csod{D} 
- 2 \Phi_{21} \sod{A} \sod{B} \csid{(C} \csod{D)} \ .\label{Riccispin}
\EEA
Following the~convention established in \cite{penrid}, we say that
$\eta$ is a weighted quantity (a scalar, a spinor, a tensor, or
an operator) of type $\{p,q\}$ if for every non-vanishing scalar field
$\lambda$ a transformation of the~form
\[
\sou{A}\mapsto \lambda \sou{A}\ ,\quad \siu{A} \mapsto \lambda^{-1}
\siu{A}\ ,\]
representing a boost in $l^\na$--$n^\na$ plane and a spatial
rotation in $m^\na$--${\bar m}^\na$ plane,
transforms $\eta$ according to
\[
\eta\mapsto \lambda^p \bar{\lambda}^q \eta\ .\]
The~boost weight, $b$,
of a weighted quantity is defined by
\[
b=\frac{1}{2}\,(p+q)\ .\]

Directional derivatives are defined by
\BEAH
D&= l^{\na} \nabla_{\na} = \sou{A} \csou{A} \dsd{A}{A} \ , \quad 
\d &= m^{\na} \nabla_{\na} = \ \sou{A} \csiu{A} \dsd{A}{A} \ , \\
D' &= n^{\na} \nabla_{\na} = \siu{A} \csiu{A} \dsd{A}{A}\ , \quad
\d' &= {\bar m}^{\na} \nabla_{\na} = \siu{A} \csou{A} \dsd{A}{A} 
\EEAH
and thus
\BE
\nabla^\na\vs \dsu{A}{A} = \siu{A} \csiu{A} D + \sou{A} \csou{A} D' - \siu{A}
\csou{A} \d - \sou{A} \csiu{A} \d' \ . \label{covder}
\EE
In the~GHP formalism new derivative operators $\thorn$, $\thorn'$, $\eth$, $\eth'$,
which are additive and obey the~Leibniz rule, are introduced.
They act on a scalar, spinor, or tensor 
$\eta$ of type $\{p,q\}$ as follows:
\BE
\eqalign{
\thorn \eta &= (D+p\ng'+q\ncg')\eta \ , \quad 
\eth \eta = (\delta - p \nb + q \ncb')\eta \ , \\
\thorn' \eta &= (D'-p\ng-q\ncg)\eta \ , \quad
\eth' \eta = (\delta' + p \nb' - q \ncb)\eta \ . \label{D_P} }
\EE
Let us explicitly write down how the~operators $\thorn$, $\thorn'$, $\eth$, $\eth'$
act on the~basis spinors
\BEA
\fl \mm\mm\mm
\thorn \sou{A}&=-\nk\siu{A} \ , \mm
\thorn \csou{A}\ =-{\bar \nk}\csiu{A} \ , \mm
\thorn \siu{A}&=-\nt'\sou{A} \ ,\mm
\thorn \csiu{A}=-{\bar \nt'}\csou{A} \ ,\nonumber \\
\fl \mm\mm\mm
\thorn' \sou{A}&=-\nt\siu{A} \ ,\mm
\thorn' \csou{A}\ =-{\bar \nt}\csiu{A} \ ,\mm
\thorn' \siu{A}&=-\nk'\sou{A} \ ,\mm
\thorn' \csiu{A}=-{\bar \nk'}\csou{A} \ ,\nonumber \\
\fl \mm\mm\mm
\eth \sou{A}&=-\ns\siu{A} \ ,\mm
\eth \csou{A}\ =-{\bar \nr}\csiu{A} \ ,\mm
\eth \siu{A}&=-\nr'\sou{A} \ ,\mm
\eth \csiu{A}=-{\bar \ns'}\csou{A} \ ,\label{derbasis} \\
\fl \mm\mm\mm
\eth' \sou{A}&=-\nr\siu{A} \ ,\mm
\eth' \csou{A}\ =-{\bar \ns}\csiu{A} \ ,\mm
\eth' \siu{A}&=-\ns'\sou{A} \ ,\mm
\eth' \csiu{A}=-{\bar \nr'}\csou{A} \ . \nonumber
\EEA
The~types and
boost-weights of various weighted quantities encountered in 
the~GHP formalism are summarized in Table 1.

\begin{table}[htbp]
\[
\begin{array}{l|c|r|c||l|c|c|c}
& p & q & b & & p & q & b \\
\hline
\vbox to 12pt{} \sou{A} & 1 & 0 & \frac{1}{2} &
\siu{A} & -1& \phneg 0 & -\frac{1}{2}\\
\hline
\nk & 3&1 & 2 & 
\nk' & -3&-1 & -2\\
\ns & 3&-1 & 1 &
\ns' & -3&\phneg 1 & -1\\
\nr & 1&1 & 1 &
\nr' & -1&-1 & -1\\
\nt & 1&-1 & 0 &
\nt' & -1&\phneg 1 &\phneg 0\\
\hline
\thorn & 1&1 & 1 &
\thorn' & -1&-1 & -1\\
\eth & 1&-1 & 0 &
\eth' & -1&\phneg 1 &\phneg 0\\
\hline
\Psi_r & 4-2r&0 & 2-r &
\Phi_{rt} & 2-2r & 2-2t & 2-r-t \\
&&&& \Lambda & 0 & 0 & \phneg 0 \\
\end{array}
\]
\caption{Boost weights of weighted quantities}
\label{tab:boost}
\end{table}

Henceforth we shall assume that conditions (A) and (B) of
Theorem~\ref{th:main} hold, and by implication that equations
\eqref{eq:Kundt}, \eqref{eq:psivanish}, \eqref{eq:phivanish},
\eqref{eq:ricvanish} hold also. Without loss of generality we also
assume that $\sou{A}$ and $\siu{A}$ are parallely propagated along
$l^{\na}$. Analytically, this condition takes the~form of
the~following two additional relations:
\begin{equation}
\label{eq:pwlog}
\ng'=0\ , 
\quad \nt'=0\ .
\end{equation}
Assumptions (A), (B) and conditions \eqref{eq:pwlog} greatly simplify
the~form of the~spin-coefficient equations, the~Bianchi and
the~commutators identities \cite{penrid}. Most of these
relations assume the~form $0=0$. Some of the non-trivial ~relations 
are as follows:
\begin{eqnarray}
\thorn\nt &=& 0\ , \label{eq:sct}\\
\thorn\ns' &=& 0\ , \label{eq:scs}\\
\thorn\nr' &=& 0\ , \label{eq:scr}\\
\thorn\nk' &=& \bar{\nt}\nr' + \nt\ns' -\Psi_3-\Phi_{21}\ , \label{eq:sck}\\
\thorn \Phi_{21} &=& 0\ , \label{eq:bianchi21} \\
\thorn \Psi_3 &=& 0\ , \label{eq:bianchi3} \\
\thorn \Phi_{22} &=& \eth \Phi_{21}+(\eth -2\nt)\Psi_3\ , \label{eq:bianchi22}\\
\thorn \Psi_4 &=& \eth' \Psi_3 + (\eth'-2\bar{\nt})\Phi_{21}\ ,
\label{eq:bianchi4} \\
\thorn\thorn'-\thorn'\thorn &=& \bar{\nt} \eth +
\nt\eth' 
\ , \label{eq:com1} \\
\thorn\eth-\eth\thorn &=& 0\ . \label{eq:com2}
\end{eqnarray}
Extending an idea introduced in \cite{typN}, we make the~following key
definition.
\begin{definition}
\label{balancedscalar}
We shall say that a weighted scalar $\eta$ with the~boost-weight $b$ is
{\em balanced} if
\BEA
&\thorn^{-b}\eta =0 \mm \mbox{for}\mm b<0\nonumber\\
\mbox{and}\mm\mm\mm & \mm\mm\eta = 0 \mm \mbox{for}\mm b\geq 0\ . \nonumber
\EEA
\end{definition}
We can now prove the~following.

\begin{lemma}
\label{lem:bal2}
If $\eta$ is a balanced scalar then $\bar{\eta}$ is also balanced.
\end{lemma}
\begin{proof}
By definition, a weighted scalar $\eta$ of type $\{p,q\}$ is changed
by complex conjugation to a weighted scalar $\bar{\eta}$ of type
$\{q,p\}$. The~boost weight, however, remains unchanged. Let us
also recall that
\[
\bar{\thorn}=\thorn\ \]
and hence that
\[
\thorn^{-b}{\bar{\eta}} = \overline{\thorn^{-b}\eta} = 0\ ,\]
as desired.
\end{proof}

\begin{lemma}
\label{lem:bal1}
If $\eta$ is a balanced scalar then 
\begin{eqnarray}
\label{eq:bal1}
&&\nt\eta,\; \nr'\eta,\; \ns'\eta,\; \nk'\eta,\; \\
\label{eq:bal2}
&&\thorn\eta,\; 
\eth\eta,\; \eth'\eta,\;\thorn'\eta
\end{eqnarray}
are all balanced as well.
\end{lemma}
\begin{proof}
Let $b$ be the boost-weight of a balanced scalar $\eta$. 
From Table~\ref{tab:boost} we see that the~scalars listed in \eqref{eq:bal1}
have boost-weights $b$, $b-1$, $b-1$, $b-2$, respectively. Hence, it
suffices to show that the following quantities are all zero:
\[
\thorn^{-b} (\nt\eta),\quad
\thorn^{1-b} (\nr'\eta),\quad
\thorn^{1-b} (\ns'\eta),\quad
\thorn^{2-b} (\nk'\eta).
\]
This follows from the~Leibniz rule and from equations
\eqref{eq:sct}, \eqref{eq:scr}, \eqref{eq:scs}, \eqref{eq:sck},
\eqref{eq:bianchi21}, and \eqref{eq:bianchi3}.

Next we show that the scalars in \eqref{eq:bal2} are balanced as
well. These scalars have boost weights $b+1$, $b$, $b$, $b-1$,
respectively. Hence, it suffices to show that the~following scalars are all
zero:
\[
\thorn^{-1-b}(\thorn \eta),\quad \thorn^{-b}(\eth \eta),\quad
\thorn^{-b}(\eth' \eta),\quad \thorn^{1-b}(\thorn' \eta).
\]
The vanishing of the~first quantity follows immediately from 
Definition~\ref{balancedscalar}. Using the~commutator relation
\eqref{eq:com2} we have
\begin{equation}
\label{eq:ethrel}
\thorn^{-b}\eth \eta = \eth\thorn^{-b}\eta = 0\ , 
\end{equation}
as desired. Vanishing of the~quantity involving $\eth'$ follows by considering
the~complex-conjugate of \eqref{eq:ethrel} and using the~relation $\bar{ \eth}=\eth'$
and Lemma~\ref{lem:bal2}.

To show that the~quantity involving $\thorn'$ vanishes, we
employ \eqref{eq:sct}, \eqref{eq:com1}, and \eqref{eq:ethrel}
to obtain
\begin{eqnarray*}
\thorn^{1-b}(\thorn' \eta) &=& \thorn^{-b}(\thorn' \thorn\eta) +
\bar{\nt}(\thorn^{-b}\eth\eta) + \nt( \thorn^{-b} \eth'\eta )
=\thorn^{-b}(\thorn' \thorn\eta)\ .
\end{eqnarray*}
We now proceed inductively and conclude that
\[
\thorn^{1-b} \thorn' \eta = \thorn' \thorn^{1-b}\eta =0\ .\]
\end{proof}

\begin{lemma}
\label{lem:sum}
If $\eta_1$, $\eta_2$ are balanced scalars both of type $\{ p,q\}$
then $\eta_1+\eta_2$ is a balanced scalar of type $\{ p,q\}$ as well.
\end{lemma}
\begin{proof}
The sum $\eta_1+\eta_2$ satisfies
\BEAH 
&&\eta_1+\eta_2 \mapsto \lambda^p \bar{\lambda}^q (\eta_1+\eta_2 )\ ,\\
&&\thorn^{-b}(\eta_1+\eta_2)=\thorn^{-b}\eta_1+\thorn^{-b}\eta_2=0\ 
\EEAH
and thus it is a balanced scalar of type $\{ p,q\}$.
\end{proof}

\begin{lemma}
\label{lem:bal3}
If $\eta_1$, $\eta_2$ are balanced scalars then $\eta_1\eta_2$ is also
balanced.
\end{lemma}
\begin{proof}
Let $b_1, b_2$ be the~respective boost weights. Boost-weights are
additive and hence the~boost-weight of the~product is $b_1+b_2$.
Setting $n=-b_1-b_2$ and applying the~Leibniz rule gives
\[
\thorn^n (\eta_1\eta_2) = \sum_{i=0}^n
\left({n\atop i}\right) \thorn^i(\eta_1)\, \thorn^{n-i}(\eta_2)\ .
\]
For $-b_1\leq i\leq n$, the~factor $\thorn^i(\eta_1)$ vanishes.
For $0\leq i\leq-1-b_1$, we have $n-i > -b_2$ and hence the~other
factor vanishes. Therefore the~entire sum vanishes.
\end{proof}

\begin{definition}
A balanced spinor is a weighted spinor of type $\{ 0,0\}$ 
whose components are all balanced
scalars.
\end{definition}

\begin{lemma}
\label{lem:prodspin} 
If ${\cal S}_1$, ${\cal S}_2$ are balanced spinors then 
${\cal S}_1{\cal S}_2$ is also a balanced spinor.
\end{lemma}
\begin{proof}
The~product ${\cal S}_1{\cal S}_2$ is a weighted spinor
of type $\{ 0,0\}$ and its components are balanced scalars
thanks to Lemma~\ref{lem:bal3}.
\end{proof}

\begin{lemma}
\label{lem:covder} 
A covariant derivative of an arbitrary order of a balanced spinor ${\cal S}$ 
is again a balanced spinor.
\end{lemma}
\begin{proof}
Applying the covariant derivative (\ref{covder}) to a balanced
spinor ${\cal S}$,
we obtain
\[
\dsu{A}{A} {\cal S} = \siu{A} \csiu{A} \thorn {\cal S}
+ \sou{A} \csou{A} \thorn' {\cal S}- \siu{A}
\csou{A} \eth {\cal S}- \sou{A} \csiu{A} \eth' {\cal S} \ .\ 
\]
From Table~\ref{tab:boost}, it follows that $\dsu{A}{A} {\cal S} $ is again a
weighted spinor of type $\{ 0,0\}$ and its components are
balanced scalars due to (\ref{derbasis}) and Lemmas~\ref{lem:bal2}, \ref{lem:bal1},
and \ref{lem:sum}.
\end{proof}

\begin{lemma}
\label{lem:contrac} 
A scalar constructed as a contraction of a balanced spinor
is equal to zero.
\end{lemma}
\begin{proof}
A scalar constructed as a contraction of a balanced spinor also has
zero boost-weight, and therefore vanishes by Definition
\ref{balancedscalar}.

Let us explain more intuitively how this works. A balanced spinor
has the~form $\sum C_i B_i$ where $C_i$ are balanced scalars and
$B_i$ are the basis spinors (products of $\sou{A}$s, $\siu{A}$s,
$\csou{A}$s, and $\csiu{A}$s). Since the~boost-weight of each $C_i$
is negative and the~boost-weight of each $C_i B_i$ is zero it
follows that the~boost-weight of each $B_i$ is positive, i.e. there
are more $\sou{A}$s and $\csou{A}$s then $\siu{A}$s and $\csiu{A}$s
in $B_i$. As a consequence of (\ref{oo_ii}) a full contraction of
each $B_i$ vanishes. In a nutshell: all scalars constructed as a
contraction of a balanced spinor vanish because each term contains
more $o$'s than $\iota$'s.

\end{proof}

We are now ready to prove that the~conditions (A) and (B) of 
Theorem~\ref{th:main} are sufficient for vanishing of all curvature
invariants. 

\begin{proof}
From Table~\ref{tab:boost}
and Eqs.~(\ref{eq:bianchi21}), (\ref{eq:bianchi3}),
(\ref{eq:bianchi22}), and (\ref{eq:bianchi4}) it follows that
the~Weyl spinor (\ref{Weylspin}) and the~Ricci spinor (\ref{Riccispin})
and their complex conjugates (Lemma~\ref{lem:bal2}) are balanced spinors.
Their products and covariant derivatives of arbitrary orders
are balanced spinors as well (Lemmas~\ref{lem:prodspin}, \ref{lem:covder}).

Finally, due to Lemma~\ref{lem:contrac} and 
Eqs.~(\ref{rozpR})--(\ref{sbsRic}) all curvature invariants
constructed from the~Riemann tensor and its covariant derivatives
of arbitrary order vanish.
\end{proof}

\section{Necessity of the conditions}
\label{sect:necessary}

In this section we consider a spacetime with vanishing curvature
invariants and prove that this spacetime satisfies the~conditions
listed in Theorem~\ref{th:main}. The~Ricci scalar, being a curvature
invariant, must vanish. To prove the~other conditions, we consider
various zeroth, first, and second order invariants formed from
the~Weyl and the~Ricci spinors, as well as the~Newmann-Penrose
equations and the~Bianchi identities.

In the~following we will employ these Newmann-Penrose equations 
\BE
\eqalign{
\thorn \nr - \eth' \nk &= \nr^2 + \ns \ncs - \nck \nt -\nk \nt'+ \Phi_{00}\ ,
\\
\eth \nr - \eth' \ns &=\nt (\nr - \ncr) + \nk (\ncr' - \nr' )-\Psi_1 + \Phi_{01}
\label{rceNP_1,2}}
\EE 
and the~Bianchi identities
\BEA
\fl&&\mm\thorn \Psi_3 - \eth' \Psi_2 - \thorn \Phi_{21} + \eth \Phi_{20} - 2 \eth' \Lambda
= 2 \ns' \Psi_1 - 3 \nt' \Psi_2 + 2 \nr \Psi_3 - \nk \Psi_4 \nonumber\\
\fl&&\mm\mm\mm - 2 \nr' \Phi_{10}
+ 2 \nt' \Phi_{11} + \nct' \Phi_{20} - 2 \ncr \Phi_{21} + \nck \Phi_{22} \ ,
\nonumber\\
\fl&&\mm\thorn \Psi_4 - \eth' \Psi_3 + \thorn' \Phi_{20} - \eth' \Phi_{21} 
= 3 \ns' \Psi_2 - 4 \nt' \Psi_3 + \nr \Psi_4 \nonumber \\
\fl&&\mm\mm\mm - 2\nk' \Phi_{10} + 2 \ns'\Phi_{11} + \ncr' \Phi_{20}
-2\nct \Phi_{21} + \ncs \Phi_{22} \ ,\label{rceB_1-4}\\ 
\fl&&\mm\thorn \Phi_{22} + \thorn' \Phi_{11} - \eth \Phi_{21} - \eth' \Phi_{12} + 3 \thorn' \Lambda
= (\nr+\ncr) \Phi_{22} + 2 (\nr' + \ncr') \Phi_{11} \nonumber \\
\fl&&\mm\mm\mm -(\nt + 2 \nct') \Phi_{21} - (2 \nt' + \nct)
\Phi_{12} - \nck' \Phi_{10} - \nk' \Phi_{01} + \ns' \Phi_{02} + \ncs' \Phi_{20} \ ,
\nonumber\\ 
\fl&&\mm\thorn' \Psi_2 - \eth \Psi_3 + \thorn \Phi_{22} - \eth \Phi_{21} + 2 \thorn' \Lambda
= \ns \Psi_4 - 2 \nt \Psi_3 + 3 \nr' \Psi_2 - 2 \nk' \Psi_1 \nonumber \\
\fl&&\mm\mm\mm + \ncr \Phi_{22} -
2 \nct' \Phi_{21} - 2 \nt' \Phi_{12} + 2 \nr' \Phi_{11} + \ncs' \Phi_{20}\ .
\nonumber 
\EEA

First, we consider the~well-known invariants
\BE
I=\ospdu{\Psi}{AB}{CD} \ospdu{\Psi}{CD}{AB} \ ,\mm 
J=\ospdu{\Psi}{AB}{CD} \ospdu{\Psi}{CD}{EF} \ospdu{\Psi}{EF}{AB} \ .
\label{invIJ}
\EE
It is generally known that these invariants vanish if and only if
the~Petrov type is III, N, or 0. In the~following we choose the~spinor
basis $\sou{A}$ and $\siu{A}$ in such a way that for the~Petrov types III and N,
$\sou{A}$ is the~multiple eigenspinor of the~Weyl spinor.
Thus the~condition (\ref{eq:psivanish}) is satisfied.

We consider the~three Petrov types case by case.

\begin{itemize}

\item[a)] Petrov type N:

\[
\Psi_0=\Psi_1=\Psi_2=\Psi_3=0.\]

Demanding that the~following invariant 
\BE
\fl
{\cal I}_1= \dsdu{D}{E} \ospu{\Psi}{{A B C D}} \dsdu{C}{D} \ospd{\Psi}{{A B K L}} 
\dsud{L}{K} \ospu{\bar \Psi}{{\dot R \dot S \dot T \dot K}} \dsud{K}{T} 
\ospd{\bar \Psi}{{\dot R \dot S \dot D \dot E}} 
= (2\Psi_4 \bar \Psi_4 \nk \nck)^2 \label{Inv_1}
\EE
vanishes we obtain
\BE
\nk = 0 \ . \label{rcekappa}
\EE
In further calculations we assume that (\ref{rcekappa}) is valid.

Vanishing of another invariant
\BE
{\cal I}_2= 
K^{F \dot F E \dot E}_{\ \ \ \ \ \ \ M \dot M L \dot L} 
{\bar K}^{M \dot M L \dot L}_{\ \ \ \ \ \ \ \ F \dot F E \dot E}
=(24 \Psi_4 {\bar \Psi_4})^2 (\rho \ncr+\ns \ncs)^4\ , \label{Inv_2} 
\EE
where
\BE
K^{F \dot F E \dot E}_{\ \ \ \ \ \ \ M \dot M L \dot L} 
= \dsu{F}{F} \dsu{E}{E} \ospu{\Psi}{{A B C D}} 
\dsd{M}{M} \dsd{L}{L} \ospd{\Psi}{A B C D}\ , \label{K}
\EE
implies that
\BE
\sigma=\rho=0 \label{rcesigmarho}
\EE
and therefore the~condition (\ref{eq:Kundt}), i.e.
the condition (A) of Theorem 1, holds.

Substituting (\ref{rcekappa}) and (\ref{rcesigmarho}) into
Eqs.~(\ref{rceNP_1,2}) we get
\BE
\Phi_{00}=\Phi_{01}=\Phi_{10}=0\ .\label{phi_0001}
\EE

And finally from the vanishing of the~invariant
\BE
\label{Isest}
\fl\mm\mm \ospu{\Phi}{AB\tA\tB} \ospd{\Phi}{AB\tA\tB}
=4{\Phi_{11}}^2+2\Phi_{02}\Phi_{20}+2\Phi_{00}\Phi_{22}
-4\Phi_{10}\Phi_{12}-4\Phi_{01}\Phi_{21}
\EE
using (\ref{phi_0001}) it follows
\[
\Phi_{11}=\Phi_{02}= {\Phi}_{20}=0\ 
\]
and thus the~condition (\ref{eq:phivanish}), i.e. the~condition (B)
of Theorem 1, is also satisfied.


\item[b)] Petrov type III:

Providing that the~Weyl spinor $\ospd{\Psi}{ABCD} $ is of Petrov type III,
we can construct another spinor $\ospd{{\tilde\Psi}}{ABCD} $ 
\BE
\ospd{{\tilde\Psi}}{ABCD} 
=\ospd{\Psi}{ABEF}\ospdu{\Psi}{CD}{\ EF}
=-2{\Psi_3}^2 \sod{A} \sod{B} \sod{C} \sod{D} \label{psi3}
\EE
which is of Petrov type N.
Now we can construct analogical curvature invariants from 
$\ospd{{\tilde\Psi}}{ABCD} $ as we
did from $\ospd{{\Psi}}{ABCD} $ for type N and again conclude that $\nk = \ns = \nr =0$
and $\Phi_{00}=\Phi_{01}= \Phi_{02}=\Phi_{11}=0 $
for metrics with all curvature invariants vanishing.

\item[c)] Petrov type 0:

Recall that the~totally symmetric Pleba\' nski spinor is defined by
\BE
\ospd{\chi}{ABCD} = \ospdu{\Phi}{(AB}{\dot{C}\dot{D}} 
\ospd{\Phi}{CD)\dot{C}\dot{D}}\ . \label{Pleb} 
\EE
Its components are
\BE
\eqalign{
\chi_0&=2 ( \Phi_{00} \Phi_{02} -{\Phi_{01}}^2)\ , \\ 
\chi_1&=\Phi_{00} \Phi_{12}+\Phi_{10}\Phi_{02}-2\Phi_{11}\Phi_{01}\ , 
\\
\chi_2&= {\scriptstyle{\frac{1}{3}}} (\Phi_{00} \Phi_{22} - 4 {\Phi_{11}}^2+\Phi_{02} \Phi_{20}
+4 \Phi_{10} \Phi_{12} - 2 \Phi_{01} \Phi_{21} ) \ ,\\
\chi_3&=\Phi_{22} \Phi_{10} + \Phi_{12} \Phi_{20} - 2 \Phi_{11} \Phi_{21}\ , 
\\
\chi_4&=2(\Phi_{22} \Phi_{20} - {\Phi_{21} }^2 ) \ .\label{Pleb_0-4}}
\EE

In analogy with the~Petrov classification of the~Weyl tensor, it is possible
to define the~Pleba\'nski-Petrov type (PP-type) of the~Pleba\'nski spinor \cite{Pleb}.
Thus vanishing of curvature invariants analogous to $I$ and $J$
(\ref{invIJ}) constructed from
the~Pleba\'nski spinor implies that the~PP-type
is III, N, or O.

For the~PP-types III and N we can argue as we did for the~cases of the~Petrov types
III and N and conclude that $\nk = \ns = \nr =0$
and $\Phi_{00}=\Phi_{01}= \Phi_{02}=\Phi_{11}=0 $.
Substituting these results into (\ref{Pleb_0-4}) 
we obtain $\chi_0=\chi_1=\chi_2=\chi_3=0$ and thus the~PP-type III is excluded.

It remains to consider the~PP-type 0 case.

Without loss of generality we can choose a null tetrad so that
\[
\Phi_{00}=0 \ .
\]
Then $\chi_0=0$ in (\ref{Pleb_0-4}) implies
\[
\Phi_{01}=\Phi_{10}=0 \ .
\]
Hence, the~vanishing of the~invariant \eqref{Isest} gives
\[
\Phi_{11}=\Phi_{02}=\Phi_{20}=0\ .
\]
Demanding $\chi_4 =0$ in (\ref{Pleb_0-4}) we get
\[
\Phi_{12}=\Phi_{21}=0 \ .
\]
Thus the~only non-vanishing component of the~Ricci spinor 
is $\Phi_{22}$. The~Bianchi identities (\ref{rceB_1-4}) 
take the form
\[
\nck\Phi_{22}=0\ ,\mm
\ncs\Phi_{22}=0\ ,\mm
\thorn\Phi_{22}=(\nr+\ncr)\Phi_{22}\ ,\mm
\thorn\Phi_{22}=\ncr\Phi_{22}\ .\mm
\]
which implies
\[
\nk=\ns=\nr=0 \ .
\]
\end{itemize}

Up to now we have proven that spacetimes with vanishing invariants constructed
from the~Weyl and the~Ricci {\it spinors} and their arbitrary derivatives satisfy
the~conditions (A) and (B) of Theorem 1. Invariants constructed from
the~Riemann {\it tensor} and its derivatives are combinations of corresponding
spinorial invariants. Thus one could argue that there might exist a very special 
class of spacetimes for which
all tensorial invariants vanish even though there exist
nonzero spinorial invariants. To prove that this does not happen
we now construct several tensorial curvature invariants. They
may also be useful for computer-aided classification of spacetimes.

The curvature invariants shown in (\ref{invIJ}) can be given as
\BEA
&& 
C_{\na \nb}^{\ \ \ \ng \d} C_{\ng \d}^{\ \ \ \na \nb} 
- i {C^{*}}_{\na \nb}^{\ \ \ \ng \d} C_{\ng \d}^{\ \ \ \na \nb} \ , \nonumber\\
&& 
C_{\na \nb}^{\ \ \ \ng \d} C_{\ng \d}^{\ \ \ \ne \phi} 
C_{\ne \phi}^{\ \ \ \na \nb} 
- i C_{\na \nb}^{\ \ \ \ng \d} {C^{*}}_{\ng \d}^{\ \ \ \ne \phi} 
C_{\ne \phi}^{\ \ \ \na \nb} \ ,\nonumber
\EEA
where 
\[
{C^{*}}_{\na \nb}^{\ \ \ \ng \d} = \pul 
\ne_{\na \nb \ne \phi} C^{\ne \phi}_{\ \ \ \ng \d}
\]
denotes the dual of the~Weyl tensor. Their vanishing implies that
the~Petrov type is III, N, or O.

\begin{itemize}

\item[a)] Petrov type N:

From the vanishing of 
\BE
I_1=C^{\na \nb \ng \d ; \ne} C_{\na \nb \nk \nl ; \ng} C^{\nr \ns \nt \nk ; \nl}
C_{\nr \ns \d \ne ; \nt} =8 {\cal I}_1 
= 2 (4 \Psi_4 \bar \Psi_4 \nk \nck)^2
\EE
$\nk = 0$ follows.

To obtain $\ns=\nr=0$ we have to construct the tensorial invariants $I_2$, $I_3$, $I_4$:
\BE
\fl
I_2=C^{\na \nb \ng \d ; \ne \phi} C_{\na \nm \ng \nn ; \ne \phi} C^{\nl \nm \nr \nn ; \ns \nt}
C_{\nl\nb\nr\d;\ns\nt} =4{\cal I}_2 +2{\cal I}_3 \bar{{\cal I}}_3
+4 ({\cal I}_4 +\bar{{\cal I}}_4 )\ ,
\EE
where
\BEA
\fl
{\cal I}_3 =
K^{F \dot F E \dot E}_{\ \ \ \ \ \ \ F \dot F E \dot E} 
=6 (4 \rho \sigma \Psi_4)^2\ , \\ 
\fl {\cal I}_4 = \dsu{F}{F} \dsu{E}{E} \ospu{\Psi}{{ABCD}} 
\dsd{F}{F} \dsd{E}{E} \ospd{\Psi}{{ACMN}}
\dsu{T}{T} \dsu{S}{S} \ospu{\Psi}{{LMNR}} 
\dsd{T}{T} \dsd{S}{S} \ospd{\Psi}{{BDLR}} \nonumber \\
\lo=18 (4 \rho \sigma \Psi_4)^4\ ,
\EEA
and thus $I_2$ is equal to
\[
\fl
I_2=2^8 3^2 \left[ (\Psi_4 \bar \Psi_4)^2 (\nr \ncr + \ns \ncs)^4 
+8(\nr \ncr \ns \ncs \Psi_4 \bar \Psi_4)^2 
+8 (\nr \ns \Psi_4)^4 + 8 (\ncr \ncs \bar \Psi_4)^4 
\right] \ .
\]
$I_3$ is defined by
\BE
I_3=C^{\na \nb \ng \d ; \ne \phi} C_{\na \nb \ng \d ; \nl \nm} 
C^{\nr \ns \nt \nn ; \nl \nm} C_{\nr \ns \nt \nn ; \ne \phi} = 
16 (2 {\cal I}_2 + {\cal I}_5 + {\bar {\cal I}}_5 )\ ,
\EE
where 
\BE
{\cal I}_5 =K^{F \dot F E \dot E}_{\ \ \ \ \ \ \ M \dot M L \dot L} 
K^{M \dot M L \dot L}_{\ \ \ \ \ \ \ \ F \dot F E \dot E}
=2^{10} 3^2 (\nr \ns \Psi_4)^4\ ,
\EE
and so it takes the form
\[
I_3=2^{11} 3^2 \left[ (\Psi_4 \bar \Psi_4)^2 (\nr \ncr + \ns \ncs)^4 
+ 8 (\nr \ns \Psi_4)^4 + 8 (\ncr \ncs \bar \Psi_4)^4 
\right] \ .
\]
The curvature invariant $I_4$ is a linear combination of $I_2$ and $I_3$
\[
I_4 = 8 I_2 - I_3 = 2^{14} 3^2 (\nr \ncr \ns \ncs \Psi_4 \bar \Psi_4)^2 \ .
\]
Demanding $I_4=0$ we obtain
\[
\nr \ns = 0
\]
and then the vanishing of $I_3$ implies
\[
\nr = \ns = 0\ .
\]

As in the spinorial case, the~Newmann-Penrose equations (\ref{rceNP_1,2})
imply $\Phi_{00}=\Phi_{01}=0$, and finally the~invariant 
constructed from the~traceless Ricci tensor corresponding to (\ref{Isest})
\[
S^{\alpha\beta} S_{\alpha\beta}=4 \ospu{\Phi}{AB\tA\tB} \ospd{\Phi}{AB\tA\tB}
=4(4{\Phi_{11}}^2+2\Phi_{02}\Phi_{20})
\]
is zero if $\Phi_{11}=\Phi_{02}= \bar{\Phi}_{20}=0$.
Thus the~conditions (A) and (B) of Theorem 1 are satisfied.

\item[b)] Petrov type III:

In analogy to (\ref{psi3}), the~tensor $D_{\na \nb \ng \d}$
can be defined in terms of the~Petrov type-III Weyl tensor 
\[
D^{\na \nb \ng \d} = C^{\na \nb}_{\ \ \ \nl \nm} C^{\ng \d \nl \nm}
\]
which is traceless, has the~same symmetries as the~Weyl tensor and is of the~Petrov
type N
\[
D^{\na \nb \ng \d} \longleftrightarrow -4 {\Psi_3}^2 \sou{A} \sou{B} \sou{C} \sou{D}
\epsu{\tA \tB } \epsu{\tC \tD } -4 {\bar \Psi_3}^2 \csou{A} \csou{B} \csou{C} \csou{D}
\epsu{{A} {B} } \epsu{{C} {D} } \ .
\]
We can construct curvature invariants from $D^{\na \nb \ng \d}$ similar to those
made from $C^{\na \nb \ng \d}$ for type N and again show that their vanishing
leads to $\nk = \ns = \nr =0$
and $\Phi_{00}=\Phi_{01}= \Phi_{02}=\Phi_{11}=0 $.

\item[c)] Petrov type O:

It is possible to define the~traceless
Pleba\' nski tensor corresponding to (\ref{Pleb}) which is
endowed with the~same symmetries as the~Weyl tensor in terms of
traceless Ricci tensor $S_{\na \nb}$ (see \cite{mac})
\[
P^{\na \nb}_{\ \ \ \ng \d} = S^{[\na}_{\ \ [\ng} S^{\nb]}_{\ \ \d]} + \d^{[\na}_{\ \ [\ng}
S_{\d ] \nl} S^{\nb ] \nl} - {{{\scriptstyle{\frac{1}{6}}}}}
\d^{[\na}_{\ \ [\ng} \d^{\nb]}_{\ \ \d]}
S_{\nm \nn} S^{\nm \nn}\ .
\]
With the~Pleba\' nski tensor we can proceed in the~same way as in the~spinorial case.

\end{itemize}

\subsection {Alternative Proof}

Another way to prove necessity of the~conditions 
(\ref{eq:Kundt})--(\ref{eq:ricvanish})
for the vanishing of all curvature invariants is
to use the~result from paper \cite{joly} 
that the invariants $I_6$, $I_7$, and $I_8$ constructed from the~Ricci spinor
are equal to zero only if all four eigenvalues of the~Ricci tensor
are equal to zero. Consequently the~Segre types of the~Ricci tensor are
$\{ (31)\}$ (i.e. PP-type N with the~only non-vanishing components
$\Phi'_{12}$ and $\Phi'_{22}$ \cite{seixas}), $\{ (211)\}$ 
(i.e. PP-type O with the~only non-vanishing component $\Phi'_{22}$), 
or $\{ (1111)\}$ (i.e. vacuum). 
In non-vacuum cases, the~multiple null eigenvector $l'$ of the~Ricci tensor
may in general differ from the~repeated null vector of the~Weyl tensor $l$;
however, by demanding vanishing of the~mixed invariants $m_1$, $m_4$, $m_6$ 
\cite{CZMJMP02}
constructed from both the~Weyl and the~Ricci tensors 
we arrive at the~condition $l'=l$. Then the~Bianchi identities
for non-vanishing $\Psi_3$, $\Psi_4$, $\Phi_{12}$, and $\Phi_{22}$
imply $\nk=0$. And finally the vanishing of the~invariant (\ref{Inv_2})
for P-types III and N
results in $\nr=\ns=0$. 

\section{Local description of the~spacetimes with vanishing curvature invariants
}\label{sect:spacetimes}

Let us describe the~metric, written in an adapted coordinate form, of all of
the~spacetimes with vanishing curvature invariants (i.e. those
satisfying Theorem~\ref{th:main}).
We recall that spacetimes with vanishing curvature invariants
satisfy \eqref{eq:Kundt} (i.e., 
belong to the~Kundt class \cite{kramer,kundt,Ludwig80}), 
are of the~Petrov type III, N, or O (i.e., 
the~Weyl spinor $\ospd{\Psi}{ABCD}$ is of the~form (\ref{Weylspin})),
and the~Ricci spinor $ \ospd{\Phi}{AB \dot{C} \dot{D}} $
has the~form (\ref{Riccispin}) that corresponds to the~Ricci tensor
\begin{equation}
R_{\alpha \beta} = -2\Phi_{22} l_{\alpha} l_{\beta} + 4 \Phi_{21} l_{(\alpha} m_{\beta)} 
+ 4 \Phi_{12} l_{(\alpha} {\bar m}_{\beta)}\ . \label{Ricci}
\end{equation}
Consequently, the~Pleba\' nski spinor (\ref{Pleb}) has the~form
\BE
\ospd{\chi}{ABCD} = -2{\Phi_{21}}^2 \sod{A} \sod{B} \sod{C} \sod{D} 
\EE
and 
the~Pleba\' nski-Petrov type (PP-type) is N for $\Phi_{12}\not= 0$
or O for $\Phi_{12}=0$.
We note that for PP-type N, using a null rotation about $l^\na$ we can transform away the~Ricci component $\Phi_{22}$
and using further a boost in the~$l^\na - n^\na$ plane and a spatial 
rotation in the~$m^\na - {\bar m}^\na$ 
plane set $\Phi_{12}=\Phi_{21}=1$. For PP-type O
it is possible to set $\Phi_{22}=1$ 
by performing a boost in the~$l^\na - n^\na$ plane.

The~Ricci tensor \eqref{Ricci}
has all four eigenvalues equal to zero 
and its Segre type is $\{ (31)\}$ ($\Phi_{12}\not= 0$), $\{ (211)\}$ 
($\Phi_{12}=0$ and $\Phi_{22}\not= 0$), 
or $\{ (1111)\}$ (for vacuum $\Phi_{12}=\Phi_{22}= 0$). 
The~most physically interesting non-vacuum
case $\{ (211)\}$ corresponds to 
a pure null radiation field \cite{kramer}.
It can be shown that an electromagnetic field compatible with \eqref{Ricci}
has to be null. Other energy-momentum tensors, including
a fluid with anisotropic pressure
and heat flux, can correspond to a Ricci tensor of PP-type O.
Indeed, it is known that no energy-momentum tensor for
a spacetime corresponding to a Ricci tensor of Segre type $\{ 31\}$ 
(or its degeneracies) can satisfy the weak energy conditions (see \cite{kramer}, p 72),
and hence spacetimes of PP-type N are not regarded as physical
in classical general relativity and hence usually attention is restricted to
PP-type O models. However, for mathematical completeness we will discuss all
of the models here. In addition, in view of possible applications in high energy physics
in which the energy conditions are not necessarily satisfied, these models may have
physical applications.

The~most general form of the~Kundt metric in
adapted coordinates $u,v, \zeta,\bar{\zeta}$ \cite{kramer} is
\BE
{\rm d}s^2=2{\rm d} u [ H {\rm d}u+{\rm d} v
+ W {\rm d}\zeta
+{\bar W} 
{\rm d}{\bar \zeta}] -2 P^{-2}{\rm d}\zeta{\rm d}{\bar\zeta}\ 
,\label{dsKundt} \EE
where the~metric functions 
\[
H=H(u,v,\zeta,{\bar\zeta}),\quad W=W(u,v,\zeta,{\bar\zeta}),\quad
P=P(u,\zeta,{\bar\zeta})\] satisfy the~Einstein equations (see
\cite{kramer} and \ref{ap:kundt}). For the spacetimes considered here,
we may, without loss of generality, put $P=1$.
The~following Tables summarize the Kundt metrics for different subcases
in the~studied class. 

\begin{table} 
\begin{center}
\begin{tabular}{|c|c|c|}
\hline
$\nt$ &P-type &metric functions \\
\hline
$= 0$ & III & $W=W_0(u,\zeta,{\bar\zeta})$ \\
& & $H=vh_1(u,\zeta,{\bar\zeta})+h_0(u,\zeta,{\bar\zeta})$ \\
& & Eqs.~(\ref{eeRuu}), (\ref{eeRuz})\\
\hline
& N & $\Psi_3=0$ (\ref{psit}) \\
& & Eqs.~(\ref{eeRuu}), (\ref{eeRuz}) \\
\hline
& O & $\Psi_3=\Psi_4=0$ (\ref{psit}), (\ref{psic}) \\
& & Eqs.~(\ref{eeRuu}), (\ref{eeRuz}) \\
\hline
$\not= 0$ & III & $W=\frac{-2v}{\zeta+{\bar\zeta}}+W_0(u,{\zeta},{\bar\zeta})$ \\
& & $H=\frac{-v^2}{(\zeta+{\bar\zeta})^2}
+v h_1(u,\zeta,{\bar\zeta})
+h_0(u,\zeta,{\bar\zeta})$ \\
& & Eqs.~(\ref{eeRuu}), (\ref{eeRuz}) \\
\hline
& N& $\Psi_3=0$ (\ref{psit}) \\
& & Eqs.~(\ref{eeRuu}), (\ref{eeRuz}) \\
\hline
& O & $\Psi_3=\Psi_4=0$ (\ref{psit}), (\ref{psic}) \\
& & Eqs.~(\ref{eeRuu}), (\ref{eeRuz}) \\
\hline
\end{tabular} 
\end{center}
\caption{All spacetimes with vanishing invariants with
$\Phi_{12}\not=0$ and $\Phi_{22}\not= 0$, i.e. PP-N,
are displayed. For details and references see \ref{ap:kundt}.}
\label{tab:PP-N}
\end{table}

\begin{table}
\begin{center}
\begin{tabular}{|c|c|c|}
\hline
$\nt$ &P-type &metric functions \\ 
\hline
$= 0$ & III & $W=W_0(u,{\bar\zeta})$ \\
& & $H=\pul v(W_0,_{\bar\zeta}+{\bar W}_0,_\zeta)
+h_0(u,\zeta,{\bar\zeta})$ \\
& & $\Phi_{22}
=h_0,_{\zeta{\bar\zeta}}-\Re (W_0W_0,_{{\bar\zeta}{\bar\zeta}}
+W_0,_{u{\bar\zeta}}+{W_0,_{\bar\zeta}}^2)$ \\
\hline
& N & $W=0$ \\
& & $H=h_0(u,\zeta,{\bar\zeta})$ \\
& & $\Phi_{22}=h_0,_{\zeta{\bar\zeta}}$ \\
\hline
& O & $W=0$ \\ 
& & $H=h_{02}(u)\zeta{\bar\zeta}+ h_{01}(u)\zeta
+{\bar h}_{01}(u){\bar \zeta}+h_{00}(u) $ \\
& & $\Phi_{22}=h_{02}(u)$ \\
\hline
$\not= 0$ & III & $W=\frac{-2v}{\zeta+{\bar\zeta}}+W_0(u,{\zeta})$ \\
& & $H=\frac{-v^2}{(\zeta+{\bar\zeta})^2}
+v \frac{W_0+{\bar W}_0}{\zeta+{\bar\zeta}}
+h_0(u,\zeta,{\bar\zeta})$ \\
& & $\Phi_{22}= (\zeta+{\bar\zeta})\lvkz
\frac{h_0+W_0{\bar W}_0}{\zeta+{\bar\zeta}}\pvkz,_{\zeta{\bar\zeta}}
-W_0,_\zeta {\bar W}_0,_{\bar\zeta}$ \\
\hline
& N& $W=\frac{-2v}{\zeta+{\bar\zeta}}$ \\ 
& & $H=\frac{-v^2}{(\zeta+{\bar\zeta})^2}
+h_0(u,\zeta,{\bar\zeta})$ \\
& & $\Phi_{22}=(\zeta+{\bar\zeta})\lvkz
\frac{h_0}{\zeta+{\bar\zeta}}\pvkz,_{\zeta{\bar\zeta}}$ \\
\hline
& O & $W=\frac{-2v}{\zeta+{\bar\zeta}}$ \\
& & $H=\frac{-v^2}{(\zeta+{\bar\zeta})^2}
+h_{00}(u)[1+h_{01}(u)\zeta+{\bar h}_{01}(u){\bar\zeta}
+h_{02}(u)\zeta{\bar\zeta}](\zeta+{\bar\zeta})$ \\
& & $\Phi_{22}=h_{00}(u)h_{02}(u)(\zeta+{\bar\zeta})$ \\
\hline
\end{tabular} 
\end{center}
\caption{All spacetimes with vanishing invariants with
$\Phi_{12}=0$ and $\Phi_{22}\not= 0$, i.e. PP-O, null radiation,
are displayed. For details and references see \ref{ap:kundt}.}
\label{tab:PP-Onull}
\end{table}

\begin{table}
\begin{center}
\begin{tabular}{|c|c|c|}
\hline
$\nt$ &P-type &metric functions \\
\hline
$= 0$ & III & $W=W_0(u,{\bar\zeta})$ \\
& & $H=\pul v(W_0,_{\bar\zeta}+{\bar W}_0,_\zeta)
+h_0(u,\zeta,{\bar\zeta})$ \\
& & $h_0,_{\zeta{\bar\zeta}}=\Re (W_0W_0,_{{\bar\zeta}{\bar\zeta}}
+W_0,_{u{\bar\zeta}}+{W_0,_{\bar\zeta}}^2)$ \\
\hline
& N & $W=0$ \\
&pp-waves & $H=h_{00}(u,\zeta)+{\bar h}_{00}(u,{\bar\zeta}) $ \\
\hline
$\not= 0$ & III & $W=\frac{-2v}{\zeta+{\bar\zeta}}+W_0(u,{\zeta})$ \\
& & $H=\frac{-v^2}{(\zeta+{\bar\zeta})^2}
+v \frac{W_0+{\bar W}_0}{\zeta+{\bar\zeta}}
+h_0(u,\zeta,{\bar\zeta})$ \\
& & $ (\zeta+{\bar\zeta})\lvkz
\frac{h_0+W_0{\bar W}_0}{\zeta+{\bar\zeta}}\pvkz,_{\zeta{\bar\zeta}}
=W_0,_\zeta {\bar W}_0,_{\bar\zeta}$ \\
\hline
& N& $W=\frac{-2v}{\zeta+{\bar\zeta}}$ \\
& & $H=\frac{-v^2}{(\zeta+{\bar\zeta})^2}
+[h_{00}(u,\zeta)+{\bar h}_{00}(u,{\bar\zeta})](\zeta+{\bar\zeta})$ \\
\hline
\end{tabular}
\end{center}
\caption{All spacetimes with vanishing invariants with
$\Phi_{12}=\Phi_{22}= 0$, i.e. PP-O, vacuum,
are displayed. For details and references see \ref{ap:kundt}.}
\label{tab:PP-Ovacuum}
\end{table}

It is of interest to find the conditions for which 
the~repeated null eigenvector $l^\na$ of the~Weyl tensor is recurrent
for the~Kundt class. The~vector $l^\na$ satisfies
\[
l^\na l_\na=0,\quad {l^\na}_{;\nb} l^\nb=0,\quad {l^\na}_{;\na}=0,\mm
l_{(\na;\nb)}l^{\na;\nb}=0,\quad l_{[\na;\nb]}l^{\na;\nb}=0\]
and its covariant derivative has in general the~form
\[
l_{\na;\nb}= 
(\ng+{\bar\ng})l_\na l_\nb 
+(\nb'-{\bar\nb}) l_\na m_\nb 
+({\bar\nb}'-\nb)l_\na {\bar m}_\nb 
-{\bar\nt} m_\na l_\nb 
-\nt{\bar m}_\na l_\nb.\]
Performing a boost in 
the~$l^\na - n^\na$ plane
\[
{\tilde l}_{\na}=A l_{\na} \ , \ \ {\tilde m}_{\na} = m_{\na} \ , 
\ \ {\tilde n}_{\na} = A^{-1} n_{\na}
\]
with 
$A$ satisfying
\[
A,_\na =A({\bar\nb}-\nb'+{\bar\nt})m_\na
+A(\nb-{\bar\nb}'+\nt){\bar m}_\na\ ,
\] 
i.e. putting ${\tilde\nb}'-{\bar{\tilde\nb}}=
\nb'-{\bar \nb}-{\d'}A/A={\bar\nt}$,
${\tilde\nt}=\nt$ (see e.g. \cite{carmeli} for 
transformation properties of NP quantities) we obtain 
\BE
{\tilde l}_{\na;\nb}= 
({\tilde\ng}+{\bar{\tilde\ng}}){\tilde l}_\na {\tilde l}_\nb 
+{\bar{\nt}}({\tilde l}_\na m_\nb -{\tilde l}_\nb m_\na)
+{\nt}({\tilde l}_\na {\bar m}_\nb -{\tilde l}_\nb {\bar m}_\na)\ 
\EE
with ${\tilde l}^\na$ 
satisfying
\BE
{\pounds}_{{\tilde l}}g_{\na\nb}={\tilde l}_{\na;\nb}+{\tilde l}_{\nb;\na}=2
({\tilde\ng}+{\bar{\tilde\ng}}) {\tilde l}_\na {\tilde l}_\nb\ . \label{almKill}
\EE
This normalization is called ``an almost Killing normalization'' in \cite{ozsvath}. 
As $\nt$ cannot be transformed away by any transformation of the~tetrad
preserving the~$l$-direction and one can even show that $\nt{\bar\nt}$ 
is invariant with respect to all tetrad transformations preserving the~$l$-direction,
the~repeated null eigenvector $l^\na$ of the~Weyl tensor is proportional 
to a recurrent vector ${\tilde l}^\na$ if and only if $\nt=0$.
To summarize: all Kundt spacetimes with $\nt=0$ admit a recurrent null vector.

Finally, let us present the~relation
between quantities $L$ and $L'$ given in \cite{ozsvath} and NP-quantities
when $l^\na$ satisfies (\ref{almKill})
\[
L=\ng+{\bar\ng}\ ,\mm
L'=DL=-2\nt{\bar\nt}\ .
\]

\section{Discussion}

The~pp-wave spacetimes have a number of important physical
applications, many of which also apply to the~other spacetimes
obtained in this paper.
As mentioned earlier, pp-wave spacetimes are exact
vacuum solutions to string theory to all order in $\alpha'$, 
the~scale set by the~string tension \cite{amati}.
Horowitz and Steif \cite{HS}
generalized this result to include the~dilaton field and
antisymmetric tensor fields which are also massless fields of
string theory
using a more geometrical approach. They showed that
pp-wave metrics satisfy all other field equations that are
symmetric rank two tensors covariantly constructed
from curvature invariants and
polynomials in the~curvature and their covariant derivatives,
and since the~curvature is null all higher order corrections
to Einstein's equation constructed from higher powers of the~Riemann tensor
automatically vanish. Therefore,
all higher-order terms in
the~string equations of motion are automatically zero. Many of
the~spacetimes obtained here will have similar properties.

In addition, solutions of classical field equations for
which the~counter terms required to regularize
quantum fluctuations
vanish are of particular importance because
they offer insights into the~behaviour of the~full quantum theory.
The~coefficients of quantum corrections
to Ricci flat solutions
of Einstein's theory of gravity in four dimensions
have been calculated up to two loops. In particular,
a class of Ricci flat (vacuum) Lorentzian 4-metrics, which includes the~pp-wave spacetimes
and some special Petrov type III or N spacetimes,
have vanishing counter terms up to and including two loops.
Thus these Lorentzian metrics suffer no quantum corrections
to all loop orders \cite{gibbons}.
In view of the~vanishing of all
quantum corrections it is possible that all of 
the~metrics summarized in Tables 2 -- 4 
are of physical import and merit further investigation.

String theory in pp-wave backgrounds has been studied by many authors
\cite{amati,tseytlin}, partly in a search for a connection between
quantum gravity and gauge theory dynamics. Such string backgrounds are
technically tractable and have direct applications to the~four
dimensional conformal theories from the~point of view of a duality
between string and gauge theories. Indeed, pp-waves provide exact
solutions of string theory \cite{HS,amati} and type-IIB superstrings
in this background were shown to be exactly solvable even in of
the~presence of the~RR five-form field strength \cite{matsaev}. As a
result the~spectrum of the~theory can be explicitly obtained. This
model is expected to provide some hints for the~study of superstrings
on more general backgrounds. There is also an interesting connection
between pp-wave backgrounds and gauge field theories. It is known
that any solution of Einstein gravity admits plane-wave backgrounds in
the~Penrose limit \cite{penrose}. This was extended to solutions of
supergravities in \cite{G}. It was shown that the~super-pp-wave
background can be derived by the~Penrose limit from the $AdS_p \times
S^q$ backgrounds in \cite{BFHP}. The~Penrose limit was recognized to
be important in an exploration of the~AdS/CFT correspondence beyond
massless string modes in \cite{AdS/CFT,BMN}. Maximally supersymmetric
pp-wave backgrounds of supergravity theories in eleven- and
ten-dimensions have attracted great interests \cite{KG}.

Recently the~idea that our universe is embedded in a higher dimensional world has
received a great deal of renewed attention \cite{brane}.
Due to the~importance of branes
in understanding the
non-perturbative dynamics of string theories,
a number of classical solutions of branes in the
background of a pp-wave have been studied; in particular
a new brane-world model has been introduced in which the~bulk
solution consists of outgoing plane waves (only), which avoids the problem
that the evolution requires initial data specified in the~bulk \cite{horo}.

Finally, in \cite{scm1} an example of
non-isometric spacetimes with non-vanishing curvature scalars
which cannot be distinguished by curvature invariants was presented.
This example represents a solution of Einstein's
equation with a negative cosmological term and
a minimally coupled massless scalar field. In this paper
we have noted the~existence of a class of spacetimes
in which all of the~curvature invariants are constants
(depending on the~cosmological constant).
These results and their extensions to higher dimensions are consequently
also of physical interest.

\ack

The~authors would like to thank R.~Zalaletdinov for discussions.
A.A.C. and R.M. were supported, in
part, by research grants from N.S.E.R.C.
A.P. and V.P. would like to
thank Dalhousie University for the~hospitality while this work was
carried out.
V.P. was supported by grant GACR-202/00/P030 
and A.P. by grant GACR-202/00/P031.

\appendix
\section{The Kundt metrics with all curvature invariants vanishing}
\label{ap:kundt}

Let us present here more details on the spacetimes
with all curvature invariants vanishing, which were briefly
summarized in Section~\ref{sect:spacetimes}. 
\ref{app:ppn}, \ref{app:pponull}, and \ref{app:ppovac}
correspond to Tables \ref{tab:PP-N}, \ref{tab:PP-Onull}, 
and \ref{tab:PP-Ovacuum}, respectively.

Since all of the~spacetimes satisfying condition (A) of Theorem 1
(i.e., that satisfy \eqref{eq:Kundt}), belong to the~Kundt class we
start with the~metric given by \eqref{dsKundt} in coordinates $u,v,
\zeta,\bar{\zeta}$ \cite{kramer}, where the~null tetrad is given by
\BE
\fl
l=\partial_v\ ,\mm
n=\partial_u-(H+P^2W{\bar W})\partial_v+P^2({\bar W}\partial_\zeta+W\partial_{\bar\zeta})\ ,
\mm m=P\partial_\zeta\ .\label{Kundttetrada}
\EE
Only certain coordinate transformations and tetrad rotations can be
performed which preserve the~form of the~metric \eqref{dsKundt} and
the~null tetrad (\ref{Kundttetrada}) (see \cite{kramer}) 

\begin{eqnarray}
\label{eq:coord1}
\rlap{\hskip -3em (I)}
\zeta' &=& f(\zeta,u)\ , \\
\nonumber
P'{}^2 &=& P^2 f,_{\zeta} \bar{f},_{\bar{\zeta}} \ ,\quad W' =
\frac{W}{f,_{\zeta}}-\frac{\bar{f},_{u}}{P^2 f,_{\zeta}
\bar{f},_{\bar{\zeta}}} ,\\
\nonumber
H' &=& H - \frac{1}{f,_{\zeta} \bar{f},_{\bar{\zeta}}}
\lvkz \frac{f,_{u}\bar{f},_{u}}{P^2} + W
f,_{u}\bar{f},_{\bar{\zeta}} + \overline{W} \bar{f},_{u}f,_{\zeta}\pvkz
\ ; \\
\label{eq:coord2}
\rlap{\hskip -3em (II)}
v' &=& v+g(\zeta,\bar{\zeta},u)\ ,\\
\nonumber
P' &=&P\ , \quad W' = W-g,_{\zeta}\ ,\quad H'=H-g,_{u}\ ;\\
\label{eq:coord3}
\rlap{\hskip -3em (III)}
u'&=&h(u)\ ,\quad v'=v/h,_{u}\ , \\
\nonumber
P' &=&P\ , \quad W' = \frac{W}{h,_{u}}\ ,\quad 
H'=\frac{1}{{h,_{u}}^2 }\lvkz H+v \frac{h,_{uu}}{h,_{u}}\pvkz \ . 
\end{eqnarray}

In particular, in these coordinates it is not possible, in general, to
simultaneously simplify the forms of the Ricci spinor components in
\eqref{Ricci} in PP-types N and O by boosts and null and spatial
rotations.

In most cases it is possible to specialize the solution form by an
appropriate choice of coordinates, thereby narrowing the range of
allowed coordinate transformations. The remaining coordinate freedom
will be described below on a case by case basis.

\subsection{Pleba\' nski-Petrov type N, i.e. $\Phi_{12}\not= 0$ and $\Phi_{22}\not= 0$}
\label{app:ppn}

\begin{itemize}
\item Petrov type III 

The~functions $H$, $W$, and $P$ have to satisfy equations which follow 
from the~fact that we assume
the~Petrov types III, N, or O ($\Psi_0=\Psi_1=\Psi_2=0$) 
and have the~Ricci tensor of the~form (\ref{Ricci}).

For the~Kundt class, $\Psi_0 $ vanishes identically and
\[
\Psi_1=\pul P R_{v\zeta}=-\ctvrt PW,_{vv}=0 
\]
and thus
\BE
W,_{vv}=0\ .
\label{Wvv}
\EE
Then $\Psi_2=0={\bar \Psi}_2$ and $R=0$ reduce to
\BEA
\fl
\mm\mm\mm\mm&&\Psi_2=-{{{\scriptstyle{\frac{1}{6}}}}} 
[H,_{vv}+2(P,_\zeta P,_{\bar\zeta}-PP,_{\zeta{\bar\zeta}})
+P^2(2W,_{v{\bar\zeta}}-{\bar W},_{v\zeta})]=0\ ,\label{Hvv}\\
\fl\mm\mm\mm\mm&&W,_{v{\bar \zeta}}={\bar W},_{v \zeta}\ ,\label{Wvcz}\\
\fl\mm\mm\mm\mm&&2W,_{v{\bar \zeta}}=W,_v{\bar W},_v \ \label{WvWv}
\EEA
and
\BE
R_{\zeta{\bar\zeta}}=-2(\ln P),_{\zeta\bar\zeta}=0\ .\label{trP}
\EE
The~Gaussian curvature, 
$K=2P^2(\ln P),_{\zeta\bar\zeta}=\bigtriangleup (\ln P)$, 
of wave surfaces determined uniquely by the~spacetime geometry
is a spacetime invariant and since it vanishes for the~studied class of spacetimes 
they are characterized by plane wave surfaces \cite{kramer}.

From (\ref{trP}), using a type I coordinate transformation
\eqref{eq:coord1} we can put
\BE P=1\ .\label{P} \EE
This restricts the type I transformations to 
\BE \zeta'=e^{i \theta(u)} \zeta + f(u)\ . \label{eq:coord1a}\EE
Then, equations (\ref{Wvv}), (\ref{Wvcz}), and (\ref{WvWv}), together with
\BE
R_{\zeta\zeta}=-
W,_{v\zeta}+{{{\scriptstyle{\frac{1}{2}}}}}{W,_v}^2=0\ ,\ \label{Wvz}
\EE
after another type I coordinate transformation \eqref{eq:coord1a},
give without loss of generality \cite{kramer}
\BE
W(u,v,\zeta,{\bar\zeta})=\frac{-2v}{\zeta+{\bar\zeta}}n+W_0(u,\zeta,{\bar\zeta})\ 
\label{W}
\EE 
with $n=0$ or $1$. If $n=1$, the wave surfaces are polarized, and
consequently type I transformations are further restricted to
\[
\zeta' = \zeta+f(u)\ ,\quad \bar{f}+f =0\ . 
\]

Finally, Eqs.~(\ref{Hvv}) and 
\BE
R_{uv}=-H,_{vv}-{{{\scriptstyle{\frac{1}{2}}}}} W,_v {\bar W},_v=0\ \label{Hvvd}
\EE
are identical and have the~solution \cite{kramer}
\BE
H(u,v,\zeta,{\bar\zeta})=\frac{-v^2}{(\zeta+{\bar\zeta})^2}n+vh_1(u,\zeta,{\bar\zeta})
+h_0(u,\zeta,{\bar\zeta})\ .\label{H}
\EE

Employing (\ref{Wvv})--(\ref{WvWv}),
(\ref{P}), (\ref{Wvz}), and (\ref{Hvvd}), the~remaining Einstein equations are
\BEA
\fl
R_{uu}= 
2({\bar W}H,_{v\zeta}+WH,_{v{\bar\zeta}})
-2H,_{\zeta{\bar\zeta}}
+H,_v(W,_{\bar\zeta}+{\bar W},_\zeta) 
-(H,_\zeta {\bar W},_v +H,_{\bar\zeta}W,_v) \nonumber\\ 
\fl \ \ \ \ \ \ -HW,_v{\bar W},_v
-(W{\bar W},_{uv}+{\bar W}W,_{uv}) 
+W,_{u{\bar\zeta}}+{\bar W},_{u\zeta}
-W,_{\bar\zeta}{\bar W},_\zeta \nonumber\\ 
\fl \ \ \ \ \ \ 
+(W{\bar W},_v-{\bar W}W,_v)(-{\bar W},_\zeta +W,_{\bar\zeta})
+\pul ({W,_{\bar\zeta}}^2+{{\bar W},_\zeta}^2
+W^2 {{\bar W},_v}^2+{\bar W}^2{W,_v}^2)
\nonumber\\
\fl \ \ \ \ \ \ 
=-2[\Phi_{22}-2(W\Phi_{21}+{\bar W}\Phi_{12})]\label{eeRuu} \\
\fl R_{u\zeta}=
-H,_{v\zeta}+\pul( W,_{uv}
-W,_{\zeta{\bar\zeta}}+{\bar W},_{\zeta\zeta} 
+W,_vW,_{\bar\zeta}-{\bar W},_vW,_\zeta)
-\ctvrt W,_v (W {\bar W},_v+{\bar W} {W,_v})
\nonumber\\
\fl \ \ \ \ \ \ 
=-2\Phi_{12}\ .\label{eeRuz}
\EEA

The~NP quantities read
\BEA
\fl \nr &=\ns=\nk=\ne=0 \ ,\nonumber\\
\fl \nt &=-\nt'
=2\nb=-2\nb'=- \frac{1}{\zeta+{\bar\zeta}}n\ , \nonumber\\
\fl 
\ns' &=-\frac{2v}{(\zeta+{\bar\zeta})^2}n-{\bar W}_0,_{\bar\zeta}\ , \nonumber\\
\fl 
\nr' &=-\frac{2v}{(\zeta+{\bar\zeta})^2}n 
-\pul (W_0,_{\bar\zeta}+{\bar W}_0,_\zeta)\ , \nonumber\\
\fl 
\nk' &=\frac{6v^2}{(\zeta+{\bar\zeta})^3}n
-v\lvhz h_1,_{\bar\zeta}+2\frac{W_0+{\bar W}_0}{(\zeta+{\bar\zeta})^2}n 
-2\frac{W_0,_{\bar\zeta}+{\bar W}_0,_{\bar\zeta} }{\zeta+{\bar\zeta}}n \pvhz
-h_0,_{\bar\zeta}-( W_0{\bar W}_0),_{\bar\zeta} \ , \nonumber\\
\fl \ng &=\frac{3v}{(\zeta+{\bar\zeta})^2}n+\pul h_1
-\frac{W_0+{\bar W}_0}{\zeta+{\bar\zeta}}n 
+\ctvrt (W_0,_{\bar\zeta}-{\bar W}_0,_\zeta)
\ , 
\nonumber\\
\fl \Psi_3&=-2h_1,_{\bar\zeta}+{\bar W}_0,_{\zeta{\bar\zeta}}-W_0,_{{\bar\zeta}{\bar\zeta}}
+2\frac{W_0,_{\bar\zeta}-{\bar W}_0,_{\bar\zeta}}{\zeta+{\bar\zeta}}n 
-2\frac{W_0+{\bar W}_0}{(\zeta+{\bar\zeta})^2}n \ ,
\label{psit}\\
\fl \Psi_4 &=v\lvhz -h_1,_{{\bar\zeta}{\bar\zeta}}
+2\frac{h_1,_{\bar\zeta}-{\bar W}_0,_{\zeta{\bar\zeta}}+W_0,_{{\bar\zeta}{\bar\zeta}}}
{\zeta+{\bar\zeta}}n
+2\frac{{\bar W}_0,_{\bar\zeta}-2W_0,_{\bar\zeta}}{(\zeta+{\bar\zeta})^2}n 
+4\frac{W_0+{\bar W}_0}{(\zeta+{\bar\zeta})^3}n \pvhz
\nonumber\\\fl &\mm
+h_1{\bar W}_0,_{\bar\zeta}-h_0,_{{\bar\zeta}{\bar\zeta}}+{\bar W}_0,_{u{\bar\zeta}}
+{\bar W}_0({\bar W}_0,_{\zeta{\bar\zeta}}-W_0,_{{\bar\zeta}{\bar\zeta}})
\nonumber\\ \fl &\mm
+2\frac{h_0,_{\bar\zeta}
+{\bar W}_0(W_0,_{\bar\zeta}-{\bar W}_0,_{\bar\zeta})}{\zeta+{\bar\zeta}}n
-2\frac{h_0+W_0{\bar W}_0}{(\zeta+{\bar\zeta})^2}n \ .\label{psic}
\EEA

The remaining coordinate freedom for the case $n=1$ is 
\begin{eqnarray}
\label{eq:coord2n=1}
\rlap{\hskip -3em (I)} 
\zeta' &=& \zeta + f(u)\ ,\quad \bar{f}+ f=0\ ,\mm {W_0}' = W_0 - f,_{u}\ ,\\
\nonumber
{h_0}' &=& h_0 - f,_{u}\bar{f},_{u} +
(W_0-\overline{W}_0) f,_{u}\ ,\mm {h_1}'=h_1\ ;\\
\nonumber
\rlap{\hskip -3em (II)} 
v' &=& v+g\ ,\quad {W_0}'= W_0 -g,_{\zeta} + 2g/(\zeta+\bar{\zeta})\ ,\\
\nonumber
{h_0}' &=& h_0 - g,_{u} - g h_1 - g^2/(\zeta+\bar{\zeta})^2\ ,\quad
{h_1}' = h_1 + 2g/(\zeta+\bar{\zeta})^2\ ;\\
\nonumber
\rlap{\hskip -3em (III)}
u'&=& h(u)\ ,\mm v'=v/h,_u\ , \\
\nonumber
{W_0}' &=& W_0/h,_{u}\ ,\quad {h_0}' = h_0/{h,_{u}}^2\ ,\quad
{h_1}' = h_1/h,_{u} + h,_{uu}/{h,_{u}}^2\ .
\end{eqnarray}
One could, without loss of generality, take $h_1=0$.

The remaining coordinate freedom for the $n=0$ case is
\begin{eqnarray}
\label{eq:coord2n=0}
\rlap{\hskip -3em (I)} 
\zeta'&=&e^{i \theta(u)} \zeta + f(u)\ ,\\
\nonumber
{W_0}' &=& e^{-i \theta} W_0 + \bar{f},_{u} - i e^{-i\theta} \theta,_{u}
\,\bar{\zeta}\ ,\quad {h_1}' = h_1\ , \\
\nonumber
{h_0}' &=& h_0 - f,_{u} \bar{f},_{u} 
- i e^{i\theta} \theta,_{u} \bar{f},_{u} \zeta
+ i e^{-i\theta} \theta,_{u} f,_{u} \bar{\zeta}
- {\theta,_{u}}^2 \zeta\bar{\zeta} \\
\nonumber
&& - W_0 ( e^{-i \theta} f,_{u} + i \theta,_{u} \zeta)
- \overline{W}_0 ( e^{i \theta} \bar{f},_{u} - i \theta,_{u} \bar{\zeta})\ ;
\\
\nonumber
\rlap{\hskip -3em (II)}
v' &=& v+g\ , \\
{W_0}' &=& W_0-g,_{\zeta}\ ,\quad
{h_0}' = h_0- g,_{u} - g h_1\ ,\mm {h_1}' = h_1\ ;\nonumber\\
\rlap{\hskip -3em (III)}
\nonumber
u'&=& h(u)\ ,\mm v'=v/h,_u\ ,  \\
\nonumber
{W_0}' &=& W_0/h,_{u}\ ,\quad {h_0}' = h_0/{h,_{u}}^2\ ,\quad
{h_1}' = h_1/h,_{u} + h,_{uu}/{h,_{u}}^2\ .
\end{eqnarray}
One could therefore without loss of generality take $h_0=0$.

In general, we cannot make any further progress unless we identify a
specific source, e.g., null radiation or null electromagnetic field,
which then yields additional field equations through Eqs.~(\ref{eeRuu})
and (\ref{eeRuz}) (and, for example, the Maxwell equations).

\item Petrov type N

In this case $\Psi_3=0$ and Eq. \eqref{psit} constitutes an
additional differential equation that must be satisfied. This
equation can be integrated to obtain a more specialized form of the
metric.

\item Petrov type O

In this case $\Psi_3=\Psi_4=0$, i.e. right hand sides of Eqs.
\eqref{psit}, \eqref{psic} must vanish. These equations can be
integrated to obtain a fully specified form of the metric.

\end{itemize}

\subsection{Pleba\' nski-Petrov type O, $\Phi_{12}=0$ and $\Phi_{22}\not= 0$ 
-- pure radiation} 
\label{app:pponull}

Conformally Ricci-flat pure radiation metrics, studied in \cite{Ludwig},
all belong to this class. In fact, in \cite{Ludwig} the authors present all
pure radiation solutions belonging to Kundt's class of Petrov types
N and O for $\nt\not= 0$ and of Petrov types III, N, and O
for $\nt=0$. For pure radiation, one of the~remaining Einstein equations simply serve to
define the radiation energy-density.
For specific sources, such as a null electromagnetic field, these equations
(e.g., Eqs.~(\ref{eeRuu}) and (\ref{eeRuz})) lead to additional differential
equations. In the~case of vacuum, all solutions can be explicitly written down
(see the next subsection).

\subsubsection{$n=0$ ($\nt=0$)}

\begin{itemize}

\item Petrov type III

For $n=0$ the~Einstein equation
$R_{u\zeta}=0$ (\ref{eeRuz}) becomes 
\[
[h_1+ {{{\scriptstyle{\frac{1}{2}}}}}
(W_0,_{\bar\zeta}-{\bar W}_0,_\zeta) ],_\zeta=0\ .\]
Using a type II transformation \eqref{eq:coord2}, \eqref{eq:coord2n=0}
(see \cite{kramer} for a discussion) and (\ref{W}),
(\ref{H}),
its solution turns out to be
\BE
\eqalign{
&W=W_0(u,{\bar\zeta})\ ,\\
&H=\pul v(W_0,_{\bar\zeta}+{\bar W}_0,_\zeta)
+h_0(u,\zeta,{\bar\zeta})\ .\label{nulIIIt0_WH}}
\EE

The~metric functions are subject to the~only remaining Einstein equation (\ref{eeRuu})
\BE
\Phi_{22}
=h_0,_{\zeta{\bar\zeta}}-\Re (W_0W_0,_{{\bar\zeta}{\bar\zeta}}
+W_0,_{u{\bar\zeta}}+{W_0,_{\bar\zeta}}^2)\ .\label{nulIIIt0_F22}
\EE

The~NP quantities take the~form
\BE
\eqalign{
\fl\nr\!\! &\!\!=\ns=\nk=\ne=\nt= 
\ns'= 
\nt'
=\nb=\nb'=0 \ ,\\ 
\fl\!\!\nr'&\!\!= -\pul (W_0,_{\bar\zeta}+{\bar W}_0,_\zeta)\ ,\mm
\nk'=-\pul v W_0,_{{\bar\zeta}{\bar\zeta}}-h_0,_{\bar\zeta}
-{\bar W}_0 W_0,_{\bar\zeta} \ ,\mm
\ng= 
\pul W_0,_{\bar\zeta}\ , 
\\
\fl\!\!\Psi_3&\!\!=-2W_0,_{{\bar\zeta}{\bar\zeta}}\ ,\mm 
\Psi_4=-\pul v W_0,_{{\bar\zeta}{\bar\zeta}{\bar\zeta}}
-h_0,_{{\bar\zeta}{\bar\zeta}}
-{\bar W}_0W_0,_{{\bar\zeta}{\bar\zeta}}\ .\label{nulIIIt0_NP}}
\EE

This choice of metric form restricts the type I, II, and III transformations
\eqref{eq:coord2n=0} to four following cases
\BEA
\zeta'&=&\zeta+f(u)\ ;\nonumber\\
v'&=&v+g_1(u)\,\zeta+
{\bar g}_1(u){\bar\zeta}+g_0(u)\ ;\nonumber\\
u'&=&a_1 u+a_0\ ,\mm v'=v/a_1\ ;\nonumber\\
u'&=&h(u),\quad v'= v/h,_{u} - (h,_{uu}/{h,_{u}}^2) \zeta\bar{\zeta}\ ,\nonumber
\EEA
where $f$, $g_1$, $g_0$, and $h$ are arbitrary functions of $u$ and 
$a_1$, $a_0$ are real constants.

\item Petrov type N

For type-N spacetimes ($\Psi_3=0$ $\rightarrow$
$W_0,_{{\bar\zeta}{\bar\zeta}}=0$), 
$W_0$ can be transformed away \eqref{eq:coord2n=0}
\cite{kramer} and thus the~metric functions 
(\ref{nulIIIt0_WH}) are
\BE
W=0\ ,\mm H=h_0(u,\zeta,{\bar\zeta})\label{nulNt0_WH} 
\EE
and the~NP quantities (\ref{nulIIIt0_NP}) read
\BE
\eqalign{
\fl \nr=\ns=\nk=\ne=\nt=
\ns'=\nt'
=\nb=\nb'=
\nr'=\ng=0 \ ,\mm
\nk'=-h_0,_{\bar\zeta} \ , \\
\Psi_3=0\ ,\mm \Psi_4=-h_0,_{{\bar\zeta}{\bar\zeta}}\ .\label{nulNt0_NP}}
\EE

The~only remaining Einstein equation (\ref{nulIIIt0_F22}) 
now becomes
\BE
\Phi_{22}=h_0,_{\zeta{\bar\zeta}}\ .\label{EnullN}
\EE

The remaining coordinate freedom comes from a mixed type I and II
transformation: 
\BEA
\label{eq:coord.2N}
\zeta' &=& e^{i\theta} (\zeta+f(u))\ ,\mm
v'=v+\bar{f},_{u}\zeta+f,_{u}\bar{\zeta} +
g(u)\ ,\\
\nonumber {h_0}' &=& h_0 -g,_{u}+f,_{u} \bar{f},_{u} -
\bar{f},_{uu}\zeta - f,_{uu} \bar{\zeta}\ , 
\EEA
where $\theta$ is a
real constant, and $u$ is determined up to a affine transformation.

These spacetimes are known as generalized pp-wave solutions. In the
case of a null electromagnetic field, energy momentum tensor, Eq.~(\ref{EnullN}) 
and Maxwell's equations lead to a further differential
equation for $h_0$, whose solution is known \cite{kramer}.

\item Petrov type O

All metrics belonging to this class are 
given in \cite{Edgar} (see (12) therein).

The~condition $\Psi_4=0$ from (\ref{nulNt0_NP}) is 
$h_0,_{{\bar\zeta}{\bar\zeta}}=0$
with the~solution 
\BE
h_0=h_{02}(u)\zeta{\bar\zeta}
\label{nulOt0_h}
\EE
after a transformation \eqref{eq:coord.2N}.
The~metric functions are thus given by (\ref{nulNt0_WH}) with
(\ref{nulOt0_h}) and the~Einstein equation (\ref{EnullN}) becomes
$\Phi_{22}=h_{02}$.

The coordinates are fixed up to an 8-parameter group of
transformations:
\BEA
\zeta' &=& e^{i \theta}(\zeta + f(u))\ , \nonumber\\
\nonumber
v' &=& v/a_1+\bar{f},_{u}\zeta+f,_{u}\bar{\zeta} + \pul (f{\bar f}),_u+g_0\ , \\
\nonumber u' &=& a_1 u + a_0\ , 
\EEA where $f(u)$ is a complex-valued
solution of
$$f,_{uu} + f h_{02} = 0,$$
and $\theta$, $a_1$, $a_0$, $g_0$ are real constants.

\end{itemize}

\subsubsection{$n=1$, $\nt\not= 0$}

\begin{itemize}

\item Petrov type III

For $n=1$, the~Einstein equation
$R_{u\zeta}=0$ (\ref{eeRuz}) is 
\[
\left [h_1+ \pul (W_0,_{\bar\zeta}-\overline{W}_0,_\zeta)
-\frac{W_0+{\bar W}_0}{\zeta+{\bar\zeta}}
\right]_{,\zeta}=
-\frac{W_{0},_{\bar{\zeta}}+\overline{W}_{0},_{\zeta}}{\zeta+\bar{\zeta}}\ .
\]

Again using a type II transformations 
\eqref{eq:coord2n=1}
(as in \cite{kramer}), we obtain the~solution (\ref{W}), (\ref{H})
\BE
\eqalign{
W&=\frac{-2v}{\zeta+{\bar\zeta}}+W_0(u,{\zeta})\ ,\\
H&=\frac{-v^2}{(\zeta+{\bar\zeta})^2}
+v \frac{W_0+{\bar W}_0}{\zeta+{\bar\zeta}}
+h_0(u,\zeta,{\bar\zeta})\ .\label{nulIIIt_WH}}
\EE

The~remaining Einstein equation (\ref{eeRuu}) then reads
\BE
\Phi_{22}=
(\zeta+{\bar\zeta})\lvkz
\frac{h_0+W_0{\bar W}_0}{\zeta+{\bar\zeta}}\pvkz,_{\zeta{\bar\zeta}}
-W_0,_\zeta {\bar W}_0,_{\bar\zeta}\ .\label{nulIIIt_F22}
\EE

The~NP quantities are as follows
\BE
\eqalign{
\fl
\nr&=\ns=\nk=\ne=0 \ ,\mm
\nt=
-\nt'
=2\nb=-2\nb'=- \frac{1}{\zeta+{\bar\zeta}}\ , \\
\fl\ns'&=-\frac{2v}{(\zeta+{\bar\zeta})^2}-{\bar W}_0,_{\bar\zeta}\ , \mm
\nr'=-\frac{2v}{(\zeta+{\bar\zeta})^2} \ , \\
\fl\nk'&=\frac{6v^2}{(\zeta+{\bar\zeta})^3}
-v\lvhz\frac{W_0+{\bar W}_0}{(\zeta+{\bar\zeta})^2}
-\frac{{\bar W}_0,_{\bar\zeta}}{\zeta+{\bar\zeta}} \pvhz
-h_0,_{\bar\zeta}- W_0{\bar W}_0,_{\bar\zeta} \ , \\
\fl\ng&=\frac{3v}{(\zeta+{\bar\zeta})^2}
-\frac{1}{2}\frac{W_0+{\bar W}_0}{\zeta+{\bar\zeta}} \ , \\
\fl\Psi_3&=-\frac{4{\bar W}_0,_{\bar\zeta} }{\zeta+{\bar\zeta}} \ ,\\
\fl\Psi_4&=v\lvhz -\frac{{\bar W}_0,_{{\bar\zeta}{\bar\zeta}}}{\zeta+{\bar\zeta}}
+\frac{6{\bar W}_0,_{\bar\zeta} }{(\zeta+{\bar\zeta})^2} \pvhz
+\frac{2h_0,_{\bar\zeta}
+{\bar W}_0,_{\bar\zeta}(W_0-{\bar W}_0)}{\zeta+{\bar\zeta}}
-2\frac{h_0+W_0{\bar W}_0}{(\zeta+{\bar\zeta})^2}
\\\fl&\mm\mm 
-h_0,_{{\bar\zeta}{\bar\zeta}}+{\bar W}_0,_{u{\bar\zeta}}\ .\label{nulIIIt_NP}}
\EE

The remaining coordinate freedom is
\BE
\eqalign{
\zeta'&=\zeta+f(u)\ ,\quad \bar{f}+ f=0\ ;\\
v' &= v+(\zeta+\bar{\zeta})g(u)\ ; \\
u'&=a_1 u+a_0\ ,\mm v'=v/a_1\ ;\\
u' &= h(u),\quad
v' = \frac{v}{h,_u}- (\zeta+\bar{\zeta})^2 \frac{h,_{uu}}{2{h,_{u}}^2}
\ .\label{eq:coord.22III}}
\EE

\item Petrov type N

All type-N pure radiation metrics were found in \cite{Ludwig}.

For type-N spacetimes ($\Psi_3=0$ $\rightarrow$ ${\bar W}_0,_{{\bar\zeta}}=0$),
$W_0$ can be transformed away again using \eqref{eq:coord2n=0},
\eqref{eq:coord.22III}, and the~metric
functions (\ref{nulIIIt_WH}) take the~form
\BE
W=\frac{-2v}{\zeta+{\bar\zeta}}\ ,\mm
H=\frac{-v^2}{(\zeta+{\bar\zeta})^2}
+h_0(u,\zeta,{\bar\zeta})\ \label{nulNt_WH}
\EE
with $h_0$ satisfying (\ref{nulIIIt_F22})
\BE
\Phi_{22}
=(\zeta+{\bar\zeta})\lvkz
\frac{h_0}{\zeta+{\bar\zeta}}\pvkz,_{\zeta{\bar\zeta}}\ .
\label{EuunullN}
\EE

The~NP quantities are as follows
\BE
\eqalign{
\fl\mm\mm\nr&=\ns=\nk=\ne=0 \ ,\mm
\nt=
-\nt'
=2\nb=-2\nb'=- \frac{1}{\zeta+{\bar\zeta}}\ ,\\
\fl\mm\mm\ns'&=\nr'=-\frac{2v}{(\zeta+{\bar\zeta})^2}\ , \mm
\nk'=\frac{6v^2}{(\zeta+{\bar\zeta})^3}
-h_0,_{\bar\zeta}\ , \mm
\ng=\frac{3v}{(\zeta+{\bar\zeta})^2}
\ , \\
\fl\mm\mm\Psi_3&=0\ ,\mm
\Psi_4=
-(\zeta+{\bar\zeta})\lvkz
\frac{h_0}{\zeta+{\bar\zeta}}\pvkz,_{{\bar\zeta}{\bar\zeta}}
\ .\label{nulNt_NP}}
\EE

The remaining coordinate freedom is given by 
\BE
\eqalign{
\zeta'&=\zeta+{\rm i} f_0\ ,\quad 
u' = h(u),\quad 
v' = \frac{v}{h,_u}- (\zeta+\bar{\zeta})^2\frac{h,_{uu}}{2{h,_{u}}^2}\ , \\
\nonumber 
{h_0}' &= \frac{h_0}{{h,_{u}}^2} 
+ \frac{(\zeta+\bar{\zeta})^2}{4 {h,_{u}}^4}
(-3 {h,_{uu}}^2 + 2 h,_{u} h,_{uuu}) \label{eq:coord.22N} }
\EE
where $f_0$ is
a real constant, and $h=h(u)$ is an arbitrary real function.

\item Petrov type O

All conformally flat pure radiation metrics
(both with $\nt=0$ and $\nt\not= 0$), generalizing the
solutions found in \cite{wils} and \cite{Koutras},
were given in \cite{Edgar}.
The physical interpretation of this class of spacetimes
is discussed in \cite{gripo}.

The~equation $\Psi_4=0$ in (\ref{nulNt_NP}) 
has the~solution (see (16) in \cite{Edgar}) 
\BE
h_0=h_{00}(u)[1+h_{01}(u)\zeta+{\bar h}_{01}(u){\bar\zeta}
+h_{02}(u)\zeta{\bar\zeta}](\zeta+{\bar\zeta})\ \label{nulOt_h}
\EE
which is to be substituted into the~metric functions (\ref{nulNt_WH}).

The~Einstein equation (\ref{EuunullN}) turns to be
\[
\Phi_{22}
=(\zeta+{\bar\zeta}) h_{00}(u)h_{02}(u)\ .
\]

The~only coordinate freedom is a translation
of $\zeta$ by an imaginary constant and $u$ is determined up to a affine transformation.

Einstein-Maxwell null fields,
massless scalar fields and neutrino fields {\it do not} exist for this
class of metrics \cite{Edgar}.

\end{itemize}

\subsection{The~vacuum case, i.e. $\Phi_{12}=\Phi_{22}=0$ }
\label{app:ppovac}

The~vacuum Petrov types-III, N, and O Kundt metrics 
are reviewed in \cite{kramer} (Chap. 27.5). The form of the metric,
and the remaining coordinate freedom are as in 
\ref{app:pponull}, with the vacuum condition imposing an additional
constraint on the metric parameters.

\subsubsection{$n=0$, $\nt=0$}

\begin{itemize}

\item Petrov type III

For vacuum Petrov type-III spacetimes, the~metric 
and the~NP quantities are given by (\ref{nulIIIt0_WH})
and (\ref{nulIIIt0_NP}), respectively, where $h_0$ satisfies
the Einstein equation 
(\ref{nulIIIt0_F22})
\BE
h_0,_{\zeta{\bar\zeta}}-\Re (W_0W_0,_{{\bar\zeta}{\bar\zeta}}
+W_0,_{u{\bar\zeta}}+{W_0,_{\bar\zeta}}^2)=0\ .
\EE

Petrov \cite{petrov} found an example belonging to this class
(in different coordinates)
\BE
{\rm d}s^2=x(v-{\rm e}^x){\rm d}u^2-2{\rm d}u{\rm d}v
+{\rm e}^x({\rm d}x^2+{\rm e}^{-2u}{\rm d}z^2) \ .
\EE

\item Petrov type N -- pp waves

The metric functions (\ref{nulNt0_WH}) and the NP quantities (\ref{nulNt0_NP}) 
of vacuum Petrov type-N spacetimes 
satisfy (\ref{EnullN}) 
\BE 
h_0,_{\zeta{\bar\zeta}}=0\ , \mm\mbox{i.e.}\mm
h_0=h_{00}(u,\zeta)+{\bar h}_{00}(u,{\bar\zeta})\ .\label{vacNt0_h}
\EE

These spacetimes belong
to the class of pp-wave spacetimes (see Chap. 21.5 in \cite{kramer})
which admit a covariantly constant null vector that is
consequently also a null Killing vector.

\item Petrov type O -- flat spacetime

For flat spacetime, Eq.~(\ref{vacNt0_h}) reduces the~solution 
(\ref{nulOt0_h}) to $h_0=0$, a flat metric.

\end{itemize}

\subsubsection{$n=1$, $\nt\not= 0$}

\begin{itemize}

\item Petrov type III

For Petrov type-III vacuum spacetimes with non-vanishing $\nt$, the~remaining
Einstein equation (\ref{nulIIIt_F22}) turns out to be
\BE
( \zeta+{\bar\zeta} )\lvkz
\frac{h_0+W_0{\bar W}_0}{\zeta+{\bar\zeta} }\pvkz,_{\zeta{\bar\zeta}}
=W_0,_\zeta{\bar W}_0,_{\bar\zeta}\ .\label{eq:E-vacIII}
\EE
Its solution determines the~metric (\ref{nulIIIt_WH}) 
and the NP quantities (\ref{nulIIIt_NP}).

An example from this class, which was originally found by Kundt \cite{kundt}, with $W_0={\bar W}_0=\psi /(\zeta+{\bar\zeta})$ 
satisfying $\psi,_{\zeta{\bar\zeta}}=0$, is known (see \cite{kramer}).

\item Petrov type N

For Petrov type-N vacuum spacetimes, the~Einstein equation (\ref{eq:E-vacIII})
simplifies to 
\[
\lvkz
\frac{h_0}{\zeta+{\bar\zeta} }\pvkz,_{\zeta{\bar\zeta}}
=0\ 
\]
with the~solution 
\BE
h_0=[h_{00}(u,\zeta)+{\bar h}_{00}(u,{\bar\zeta})](\zeta+{\bar\zeta})\ .
\label{vacNt_h}
\EE
The~metric and NP quantities are then given by (\ref{nulNt_WH})
and (\ref{nulNt_NP}) with (\ref{vacNt_h}).

\item Petrov type O -- flat spacetime

For the~flat spacetime the~condition $\Psi_4=0$ in (\ref{nulNt_NP}), i.e.
\[
\lvkz
\frac{h_0}{\zeta+{\bar\zeta} }\pvkz,_{{\bar\zeta}{\bar\zeta}}
=0\ ,
\]
has the~solution
\[
h_0=h_{00}(u)[1+h_{01}(u)\zeta+{\bar h}_{01}(u){\bar\zeta}
](\zeta+{\bar\zeta})\ .
\]

\end{itemize}

{}

\end{document}